\documentclass[hyper,a4paper]{JHEP3}
\usepackage{multirow}
\pdfoutput=1
\usepackage{cite}
\usepackage{subfigure}
\usepackage{rotating}
\usepackage{amssymb}
\usepackage{amsmath}
\usepackage{blkarray}
\usepackage{multirow}
\usepackage{mathtools}
\usepackage{relsize}
\def\lesssim{\mathrel{\hbox{\rlap{\hbox{\lower5pt\hbox{$\sim$}}}\hbox{$<$}}}}
\def\gtrsim{\mathrel{\hbox{\rlap{\hbox{\lower5pt\hbox{$\sim$}}}\hbox{$>$}}}}

\catcode`\@=11
\def\lsim{\mathrel{\mathpalette\@versim<}}
\def\gsim{\mathrel{\mathpalette\@versim>}}
\def\@versim#1#2{\vcenter{\offinterlineskip
\ialign{$\m@th#1\hfil##\hfil$\crcr#2\crcr\sim\crcr } }}
\catcode`\@=12
\def\be{\begin{equation}}   %
\def\ee{\end{equation}}   %
\def\bea{\begin{eqnarray}}   %
\def\eea{\end{eqnarray}}   %

\title{Naturality vs perturbativity, $B_s$ physics, and LHC data in triplet extension of MSSM}
\author{Priyotosh Bandyopadhyay$^{1}$, Stefano Di Chiara$^{2}$, Katri Huitu$^{3}$ and Asl{\i} Sabanc{\i} Ke\c{c}eli$^{4}$\\

Department of Physics,  and Helsinki Institute of Physics,\\
P.O.Box 64 (Gustaf H\"allstr\"omin katu 2), FIN-00014 University of Helsinki, Finland\\
Email: \email{$^1$priyotosh.bandyopadhyay@helsinki.fi,$^2$stefano.dichiara@helsinki.fi, $^3$katri.huitu@helsinki.fi,
$^4$asli.sabanci@helsinki.fi}}

\abstract{
In this study we investigate the phenomenological viability of the $Y=0$ Triplet Extended Supersymmetric Standard Model (TESSM) by comparing its predictions with the current Higgs data from ATLAS, CMS, and Tevatron, as well as the measured value of the $B_s\to X_s \gamma$ branching ratio. We scan numerically the parameter space for data points generating the measured particle mass spectrum and also satisfying current direct search constraints on new particles. We require all the couplings to be perturbative up to the scale $\Lambda_{\rm UV}=10^4$ TeV, by running them with newly calculated two loop beta functions, and find that TESSM retains perturbativity as long as $\lambda$, the triplet coupling to the two Higgs doublets, is smaller than 1.34 in absolute value. For $|\lambda|\gtrsim 0.8$ we show that the fine-tuning associated to each viable data point can be greatly reduced as compared to values attainable in MSSM. We also find that for perturbatively viable data points it is possible to obtain either enhancement or suppression in $h\rightarrow \gamma \gamma$ decay rate depending mostly on the relative sign between $M_2$ and $\mu_D$. Finally, we perform a fit by taking into account 58 Higgs physics observables along with $\mathcal{B}r(B_s\to X_s \gamma)$, for which we calculate the NLO prediction within TESSM. We find that, although naturality prefers a large $|\lambda|$, the experimental data disfavors it compared to the small $|\lambda|$ region, because of the low energy observable $\mathcal{B}r(B_s\to X_s \gamma)$. We notice, though, that this situation might change with the second run of LHC at 14 TeV, in case the ATLAS or CMS results confirm, with smaller uncertainty, a large enhancement in the Higgs decay channel to diphoton, given that this scenario strongly favours a large value of $|\lambda|$.}

\begin{document}

\section{Introduction}
The discovery of the Higgs boson with a mass around 126 GeV, which has been reported by the CMS and ATLAS collaborations \cite{Higgsd1,Higgsd2}, opens up a new era in understanding the origins of the electroweak (EW) symmetry breaking. However, questions regarding the theory behind the observed spin 0 particle still need to be addressed. Even though the recent experimental results obtained in the $ZZ$ \cite{ZZ1,ATLAS:2013nma}, $WW$ \cite{WW1,ATLAS:2013wla}, $b\bar{b}$ \cite{Chatrchyan:2013zna,ATLAS:2012aha}, $\tau\tau$ \cite{Chatrchyan:2014nva,tau2}, and $\gamma\gamma$ \cite{CMS:2014ega,ATLAS:2013oma} decay channels are compatible with the Standard Model (SM), there is still room for theories beyond the SM that can accommodate more than one Higgs boson with a non-standard Higgs structure. These models are motivated by the problems in the SM such as the naturalness of the Higgs mass and lack of a dark matter candidate.

Supersymmetric models remain among the best motivated extensions of the SM. The Constrained Minimal Supersymmetric Standard Model (CMSSM) \cite{cMSSM} is a well studied model with a minimal set of parameters and a dark matter candidate. Recent studies \cite{cmssmstd,cmssmfit} have shown that, within CMSSM, it is difficult to generate a Higgs boson with mass around 126 GeV consistent with all experimental constraints from colliders as well as with the observed dark matter relic abundance and muon anomalous magnetic moment. Indeed the measured Higgs boson mass can be achieved only for large values of the CMSSM dimensional parameters, $m_0$ and $m_{1/2}$. The experimentally viable regions of parameter space result in a multi-TeV sparticle spectrum that generates a fine-tuning $<0.1\%$  \cite{cmssmfinet}. 

In general MSSM the desired Higgs mass can be achieved with the help of radiative corrections for a large mixing parameter, $A_t$, which in turn generates a large splitting between the two physical stops \cite{mssmsd}, and/or large stop soft squared masses. It was shown in \cite{mssmft} that MSSM parameter regions allowed by the experimental data require tuning smaller than $1\%$, depending on the definition of fine-tuning. Such a serious fine-tuning can be alleviated by having additional tree-level contributions to the Higgs mass, given that in MSSM the tree-level lightest Higgs is restricted to be lighter than $m_Z$, so that sizable  quantum corrections are no longer required. In order to have additional contributions to the tree-level lightest Higgs mass, one can extend the MSSM field content by adding a singlet \cite{NMSSMft} and/or a triplet \cite{Espinosa:1991wt,Espinosa:1991gr,DiChiara:2008rg,chargedH,tripletft2,Huitu:1997rr,Zhang:2008jm,Kang:2013wm} chiral superfield(s).

Another advantage of singlet and triplet extensions of MSSM concerns CP symmetry breaking. Any softly broken low energy supersymmetric theory provides general soft breaking terms with complex phases which are necessary to explain the baryon asymmetry of the universe along with the CKM matrix of the SM \cite{baryon}. However, such explicit CP violation scenarios can lead to overproduction of CP violation that is stringently constrained by electric dipole moments (EDMs) \cite{EDMs}. This overproduction problem can be naturally evaded by breaking CP symmetry spontaneously. In the case of MSSM, spontaneous CP-violation is not feasible even at higher orders because of the existing experimental bounds on the Higgs masses \cite{SCPVMSSM}. The spontaneous CP violation can be achieved in the extended models with new singlet \cite{SCPVsinglet} or triplet superfield(s) \cite{SCPVtriplet}. 

In light of fine-tuning considerations as well as the motivation of having spontaneous CP violation, here we consider the Triplet Extended Supersymmetric Standard Model (TESSM)\cite{Espinosa:1991wt,Espinosa:1991gr}. The model we consider here possesses a $Y=0$ SU(2) triplet chiral superfield along with the MSSM field content, where the extended Higgs sector generates additional tree-level contributions to the light Higgs mass and moreover may enhance the light Higgs decay rate to diphoton \cite{DiChiara:2008rg,tessm1,Delgado:2012sm,Delgado:2013zfa}. 

To assess the viability of TESSM for the current experimental data, we perform a goodness of fit analysis, by using the results from ATLAS, CMS, and Tevatron on Higgs decays to $ZZ,WW,\gamma\gamma,\tau\tau,b\bar{b}$, as well as the measured $B_s\to X_s \gamma$ branching ratio, for a total of 59 observables. Several similar fits have been performed for MSSM \cite{Arbey:2012dq,Bechtle:2012jw,Djouadi:2013uqa,Buchmueller:2013rsa} and for NMSSM \cite{Gunion:2012zd,D'Agnolo:2012mj}, but to the best of our knowledge no such goodness of fit analysis of TESSM is present in the literature. As free parameters we use Higgs coupling coefficients associated with each SM field, as well as two extra parameters that take into account the contribution of the non-SM charged and coloured particles of TESSM to the loop induced Higgs decays to diphoton and digluon, respectively. As explained later in the text, in the viable region of the TESSM parameter space the $W$ and $Z$ bosons have a SM-like coupling to the light Higgs, and, in the same region, the upper and lower components of EW SM fermion doublets have coupling coefficients which are ultimately functions only of $\tan\beta$, the ratio between the vacuum expectation value(s) (vev) of the up and down Higgs doublets. The total number of free parameters of TESSM for the fit we perform is therefore reduced to just three, plus one to fit the $\mathcal{B}r(B_s\to X_s \gamma)$ data.

An important result of the fit is that, for viable data points in the TESSM parameter space, we observe not only an enhancement of the Higgs decay to diphoton, as previously observed in \cite{DiChiara:2008rg,Delgado:2012sm,Delgado:2013zfa,Arina:2014xya}, but also a suppression of the same decay rate. This is due to the fact that we scan also negative values of mass and coupling parameters, for which the light chargino mass  and its coupling to the light Higgs can have the same sign. This, as it is the case for the top quark, produces a destructive interference between the $W$ and triplino-like chargino contributions to the Higgs decay to diphoton.

In this article we also consider the low energy observable $\mathcal{B}r(B_s\to X_s \gamma)$ to constrain the model and improve the relevance of the fit we perform. In general, the $B$ meson observables, {\it e.g.} $\mathcal{B}r(B_s\to X_s \gamma)$ and $\mathcal{B}r(B_s\to \mu^+\mu^-)$, are used to set constraints on the parameter space of the theories beyond the SM. It has been shown that, for low values of $\tan{\beta}$, the flavour bounds obtained from $\mathcal{B}r(B_s\to X_s\gamma)$ are relevant, while the constraints from $\mathcal{B}r(B_s\to \mu^+\mu^-)$ play a decisive role only for $\tan\beta \gtrsim 10$ \cite{Btomumu}. As we focus on the low  $\tan{\beta}$ region ($\lsim$10), given that the contribution of the triplet field to the Higgs mass grows as $\sin2\beta$, we study here only $\mathcal{B}r(B_s\to X_s \gamma)$. In \cite{tessm1} we already considered this constraint in the context of the lightest charged Higgs and the lightest chargino as they dominantly contribute to the decay. Here we have improved our analysis by considering the contributions from all charged Higgses and charginos at next to the leading order (NLO).

The rest of the paper is organized as follows. In the next Section, we give a brief description of the model. In Section~\ref{HmConst} we discuss the minimum of the TESSM scalar potential which leads to an extra contribution to the tree level lightest Higgs mass. In the same Section we describe the method we use to evaluate numerically the radiative corrections and find data points with a Higgs mass around 126 GeV that satisfy the current direct search limits on new particles. Section~\ref{finetuningTESSM} is devoted to the discussion on the fine-tuning associated to viable data points in TESSM. By running the dimensionless couplings with two loops beta functions, we show that there is a tension between the requirement of perturbativity at high scales and the possibility to reduce the amount fine-tuning typical for MSSM. In Section~\ref{Hphy} we consider the Higgs decay modes, especially Higgs decay to two photons for which our results partially differ from the  ones obtained previously. In Section~\ref{btsgsec} we present the results of the calculation of $\mathcal{B}r(B_s\to X_s \gamma)$ at NLO in TESSM. Section~\ref{fitsect} is dedicated to the goodness of fit analysis of TESSM considering different experimental constraints from LHC and Tevatron along with $\mathcal{B}r(B_s\to X_s \gamma)$. In Section~\ref{concsec} we finally offer our conclusions.

\section{The Model} \label{modintr}
The field content of TESSM is the same as that of the MSSM with an additional field in the adjoint of SU$(2)_L$, the triplet chiral superfield $\hat T$, with zero hypercharge ($Y=0$), where the scalar component $T$ can be written as
\be
T=\left(\begin{array}{cc}\frac{1}{\sqrt{2}} T^0 & T^+ \\T^- & -\frac{1}{\sqrt{2}}T^0\end{array}\right)\ .
\ee
The renormalizable superpontential of TESSM includes only two extra terms as compared to MSSM, given that the cubic triplet term is zero:
\be
W_{\rm TESSM}=\mu_T {\rm Tr}(\hat T \hat T) +\mu_D \hat H_d\!\cdot\! \hat H_u + \lambda \hat H_d\!\cdot\! \hat T \hat H_u + y_t \hat U \hat H_u\!\cdot\! \hat Q - y_b \hat D \hat H_d\!\cdot\! \hat Q- y_\tau \hat E \hat H_d\!\cdot\! \hat L\ ,
\label{SP}
\ee
where "$\cdot$" represents a contraction with the Levi-Civita symbol $\epsilon_{ij}$, with $\epsilon_{12}=-1$, and a hatted letter denotes the corresponding superfield. Note that the triplet field couples to the Higgs doublets through the coupling $\lambda$. The soft terms corresponding to the superpotential above and the additional soft masses can be written similarly\footnote{We use the common notation using a tilde to denote the scalar components of superfields having a SM fermion component.} as
\bea\label{softV}
V_S&=&\left[\mu_T B_T {\rm Tr}(T T) +\mu_D B_D H_d\!\cdot\! H_u + \lambda A_T H_d\!\cdot\! T H_u + y_t A_t \tilde{t}^*_R H_u\!\cdot\! \tilde{Q}_L + h.c.\right]  \nonumber\\
       & & + m_T^2 {\rm Tr}(T^\dagger T) + m_{H_u}^2 \left|H_u\right|^2  + m_{H_d}^2 \left|H_d\right|^2 + \ldots  \ ,
\eea
where we have included only the top squark cubic term, among those in common with MSSM\footnote{The neglected cubic terms are not necessary for phenomenological viability in the analysis we perform in this work.}, and wrote explicitly the squared soft mass terms only for the three scalar fields with neutral components. In the following we assume all the coefficients in the Higgs sector to be real, as to conserve CP symmetry. We moreover choose real vevs for the scalar neutral components, so as to break correctly EW symmetry SU$(2)_L\times$ U$(1)_Y$:
\be\label{vev}
\langle T^0\rangle=\frac{v_T}{\sqrt{2}}\ ,\quad\langle H_u^0\rangle=\frac{v_u}{\sqrt{2}}\ ,\quad\langle H_d^0\rangle=\frac{v_d}{\sqrt{2}}\ ,
\ee
which generate the EW gauge bosons masses
\be
m_W^2=\frac{1}{4} g_L^2 \left( v^2 + 4 v_T^2 \right)\ ,\quad m_Z^2=\frac{1}{4} \left(  g_Y^2+ g_L^2 \right) v^2\ ,\quad v^2=v_u^2+v_d^2\ .
\ee
From these masses we find that there is a non-zero tree-level contribution to the EW $\alpha_eT$ parameter \cite{Peskin:1991sw,Burgess:1993vc}:
\be
\alpha_e T=\frac{\delta m_W^2}{m_W^2}=\frac{4 v_T^2}{v^2}\ ,
\ee
with $\alpha_e$ being the fine structure constant. The measured value of the Fermi coupling $G_F$ and the upper bound on the EW parameter $T$ ($\alpha_e T\leq 0.2$ at 95\% CL) \cite{Beringer:1900zz} then impose
\be
v_w^2=v^2+4 v_T^2=\left(\rm 246~GeV\right)^2\ ,\quad v_T \lesssim 5~{\rm GeV}\ .
\ee
Such a small value of the triplet vev evidently does not allow the triplet extension to solve the MSSM $\mu$ problem. Thus, the $\mu_D$ term is defined separately in the superpotential Eq. \eqref{SP}. Given that the triplet vev can still generate small differences in the light Higgs couplings to SM particles as compared to MSSM, throughout this paper we take a small but non-zero fixed value for $v_T$:
\be
v_T= 3\sqrt{2} ~{\rm GeV}\ .
\ee
Having defined a viable EW symmetry breaking minimum, in the next Section we proceed to determine the mass spectrum of TESSM.

\section{Higgs Mass \& Direct Search Constraints} \label{HmConst}

After EW symmetry breaking, the stability conditions for the full potential are defined by
\bea
\partial_{a_i} V|_{\rm vev}&=&0\ ,\quad V=V_D+V_F+V_S\ , \quad \langle a_i\rangle = v_i \ , \quad  i=u,d,T\ ; \nonumber\\ 
 H^0_u&\equiv& \frac{1}{\sqrt{2}}\left(a_u+i b_u  \right)\ , \quad H^0_d\equiv \frac{1}{\sqrt{2}}\left(a_d+i b_d  \right)\ , \quad T^0\equiv \frac{1}{\sqrt{2}}\left(a_T+i b_T  \right)\ ,
\eea
where $V_D$ and $V_F$ are the $D$ and $F$ terms of the potential, respectively, while $V_S$ is given in Eq.~\eqref{softV}, and $a_i$ and $b_i$ are both real. The conditions above allow one to determine three of the Lagrangian free parameters:
\bea\label{stabV}
m_{H_u}^2&=&-\mu _D^2-\frac{g_Y^2+g_L^2}{8}  \left(v_u^2-v_d^2\right)+B_D \mu _D \frac{v_d}{v_u}-\frac{\lambda^2}{4}  \left(v_d^2+v_T^2\right)+\lambda  \left(\mu
   _D-\left(\frac{A_T}{2}+\mu _T\right)\frac{v_d}{v_u}\right) v_T\ ,\nonumber\\
m_{H_d}^2&=&-\mu _D^2+\frac{g_Y^2+g_L^2}{8}  \left(v_u^2-v_d^2\right)+B_D \mu _D \frac{v_u}{v_d}-\frac{\lambda^2}{4} \left(v_u^2+v_T^2\right)+\lambda  \left(\mu
   _D-\left(\frac{A_T}{2}+\mu _T\right)\frac{v_u}{v_d}\right) v_T\ ,\nonumber\\
 m_T^2&=&-\frac{\lambda ^2}{4}  \left(v_d^2+v_u^2\right)-2 \mu _T \left(B_T+2 \mu _T\right)+\lambda  \left(\mu _D \frac{v_d^2+v_u^2}{2 v_T}-
   \left(\frac{A_T}{2}+\mu _T\right) \frac{v_d v_u}{v_T}\right)     \ .
\eea
A simple condition that the remaining parameters have to satisfy for successful EW symmetry breaking is obtained by requiring the trivial vacuum at the origin to be unstable. By taking all the vevs to be zero, the requirement that one of the eigenvalues of ${\cal M}^2_{h^{0}}$, the neutral scalar squared mass matrix given in Eq.~\eqref{mns}, be negative, gives the condition
\be
B_D^2>\mu _D^2 \left(\frac{m_{H_d}^2}{\mu _D^2}+1\right) \left(\frac{m_{H_u}^2}{\mu _D^2}+1\right)\ .
\ee
When the condition above is satisfied, one can derive an important bound on the mass of the lightest neutral Higgs: given that the smallest eigenvalue of a $3\times 3$ Hermitian positive definite matrix, in this case ${\cal M}^2_{h^{0}}$, cannot be greater than the smaller eigenvalue of either of the $2\times 2$ submatrices on the diagonal, in the limit of large $B_D$ one obtains \cite{Espinosa:1991wt,Espinosa:1991gr}
\be\label{mhbnd}
m^2_{h^0_1}\leq m_Z^2 \left( \cos{2\beta} + \frac{\lambda^2}{g_Y^2+g_L^2} \sin{2\beta} \right)\ ,\quad \tan\beta=\frac{v_u}{v_d}\ .
\ee
The result in Eq.~\eqref{mhbnd} shows the main advantage and motivation of TESSM over MSSM: for $\tan\beta$ close to one and a large $\lambda$ coupling it is in principle possible in TESSM to generate the experimentally measured light Higgs mass already at tree-level \cite{DiChiara:2008rg}, which would imply no or negligible Fine-Tuning (FT) of the model. Indeed $\lambda\sim1$ and $\tan\beta\sim 1$ already saturate the bound in Eq.~\eqref{mhbnd}. Such large value of $\lambda$ in general grows nonperturbative at the GUT scale, and therefore also for TESSM, like for MSSM, radiative corrections are necessary to generate a light Higgs mass equal to 125.5~GeV \cite{Aad:2012tfa,Chatrchyan:2012ufa}.

\subsection{One Loop Potential}

The one loop contribution to the scalar masses is obtained from the Coleman-Weinberg potential \cite{Coleman:1973jx}, given by
\begin{align}\label{VCW}
V_{\rm CW}=\frac{1}{64\pi^2}{\rm STr}\left[ \mathcal{M}^4
\left(\log\frac{\mathcal{M}^2}{\mu_r^2}-\frac{3}{2}\right)\right],
\end{align}
where $\mathcal{M}^2$ are field-dependent mass matrices in which the fields are not replaced with their vevs nor the soft masses with their expressions at the EW vacuum, $\mu_r$ is the renormalization scale, and the supertrace includes a factor of $(-1)^{2J}(2J+1)$, with the spin degrees of freedom appropriately summed over. The corresponding one loop contribution to the neutral scalar mass matrix, $\Delta{\cal M}^2_{h^{0}}$, is given by \cite{Elliott:1993bs,DiChiara:2008rg}

\begin{align}
(\Delta\mathcal{M}^2_{h^0})_{ij}
&=\left.\frac{\partial^2 V_{\rm{CW}}(a)}{\partial a_i\partial a_j}\right|_{\rm{vev}}
-\frac{\delta_{ij}}{\langle a_i\rangle}\left.\frac{\partial V_{\rm{CW}}(a)}{\partial a_i}\right|_{\rm{vev}}
\label{1Lmha}\\
&=\sum\limits_{k}\frac{1}{32\pi^2}
\frac{\partial m^2_k}{\partial a_i}
\frac{\partial m^2_k}{\partial a_j}
\left.\ln\frac{m_k^2}{\mu_r^2}\right|_{\rm{vev}}
+\sum\limits_{k}\frac{1}{32\pi^2}
m^2_k\frac{\partial^2 m^2_k}{\partial a_i\partial a_j}
\left.\left(\ln\frac{m_k^2}{\mu_r^2}-1\right)\right|_{\rm{vev}}
\nonumber\\
&\quad-\sum\limits_{k}\frac{1}{32\pi^2}m^2_k
\frac{\delta_{ij}}{\langle a_i\rangle}
\frac{\partial m^2_k}{\partial a_i}
\left.\left(\ln\frac{m_k^2}{\mu_r^2}-1\right)\right|_{\rm{vev}}\ ,\quad i,j=u,d,T\ ;
\label{1Lmh}
\end{align}
where the second term in Eq.~\eqref{1Lmha} takes into account the shift in the minimization conditions, and $\{m^2_k\}$ is the set of eigenvalues of the field dependent mass matrices, which for the reader's convenience are given in the Appendix~\ref{massmapp}. Though the supertrace expressions are dropped in Eq.\eqref{1Lmh} for simplicity, the proper coefficient for each mass eigenvalue is taken into account in the calculation. Given that we include terms mixing the gauginos and higgsinos in the neutralino mass matrix, the mass matrices that enter Eq.\eqref{1Lmh} through their eigenvalues can be as large as $5\times 5$: to simplify the task of finding the one loop mass of the neutral scalars, we evaluate the derivatives in Eq.~\eqref{1Lmh} numerically at randomly assigned values for the independent parameters and for finite, though small, differentials $\Delta a_i$ around their respective vevs $v_u,v_d,v_T$, at a renormalization scale $\mu_r=m_Z$. For each randomly chosen point in the TESSM parameter space we check that, by changing the size of $\Delta a_i$ relative to $v_i$, the values of the neutral scalar masses are stable within a 0.1\% error or less.

To evaluate the phenomenological viability of TESSM we proceed by scanning randomly the parameter space for points that give the correct light Higgs mass while satisfying the constraints from direct searches of non-SM particles. The region of parameter space that we scan is defined by:
\bea\label{pscan}
&&1\leq t_{\beta }\leq 10\ ,\ 5 \,\text{GeV}\leq \left|\mu _D,\mu _T\right|\leq 2 \,\text{TeV}\ ,\ 50 \,\text{GeV}\leq \left|M_1,M_2\right|\leq 1  \,\text{TeV}\ ,\nonumber\\ 
&& \left| A_t,A_T,B_D,B_T\right|\leq 2 \,\text{TeV}\ ,\ 500 \,\text{GeV}\leq m_Q,m_{\tilde{t}},m_{\tilde{b}}\leq 2 \,\text{TeV}\ ,
\eea
with the last three being, respectively, the left- and right-handed squark squared soft masses. The value of $\lambda$ at each random point in the parameter space is determined by matching the lightest Higgs mass at one loop to 125.5~GeV: the matching is achieved by an iterative process that starts by assigning an initial random value $\left|\lambda\right|\leq 2$ to calculate the one loop contribution to the lightest Higgs mass $m_{h^0_1}^2$, solving for the value of $\lambda$ in the tree level contribution needed to match the measured light Higgs mass, using this value of $\lambda$ in place of the initial random value to calculate $m_{h^0_1}^2$, and repeating the process until $\lambda$ remains constant after the next iteration. We imposed no constraint on the sign of $\lambda$. The remaining free parameters of TESSM are of little relevance for the observables we consider in the rest of this paper (Higgs production and decay rates and $B_s\rightarrow X_s\gamma$ branching ratio), and can therefore be considered to be fixed to values consistent with the current experimental limits on new physics. Having implemented the setup outlined above, we scan randomly the parameter space defined in Eq.~\eqref{pscan} and collect 13347 points that satisfy the constraints
\bea
m_{h_1^0}=125.5\pm 0.1\, {\rm GeV}\ ;\ m_{A_{1,2}},\ m_{\chi^0_{1,2,3,4,5}}&\geq & 65\,{\rm GeV}\ ;\nonumber\\
m_{h^0_{1,2}} , m_{h^\pm_{1,2,3}}, m_{\chi^\pm_{1,2,3}}\geq 100\,{\rm GeV} \ ;\ m_{\tilde{t}_{1,2}},m_{\tilde{b}_{1,2}}&\geq & 650\,{\rm GeV}\ .
\eea
The experimental bounds \cite{Beringer:1900zz} on the mass of pseudoscalars and neutralinos are actually less tight than the ones above, but we prefer to avoid in this general study the phenomenological complicacies of invisible decays of the light Higgs, which are though relevant for dark matter \cite{Arina:2014xya}. In Section~\ref{finetuningTESSM} we impose additional, coupling dependent constraints on the heavy neutral Higgses. Before doing that, in the next Section we take up the task of studying the running of the coupling constants at high energy, and require that those couplings stay perturbative all the way up to $\Lambda_{\rm UV}$, a UV scale suitable for TESSM. This requirement, in turn, imposes a limit on the minimum amount of FT that TESSM can achieve.

\section{Perturbativity vs Fine-Tuning} 
\label{finetuningTESSM}
In the parameter space scan we allow $\lambda$ to take up absolute values larger than 1, given that these generate a light Higgs mass that can easily match 125.5~GeV already at tree-level. Such large couplings, though, can easily diverge to infinity at high scales, making the perturbative treatment of the model inconsistent. We therefore calculate the two loop beta functions for the dimensionless couplings of the superpotential and the gauge couplings ($y_t,y_b,y_{\tau },\lambda ,g_3,g_2=g_L,g_1=\sqrt{5/6}\,g_Y$), for the first time for TESSM, and run each coupling from the renormalization scale $\mu_r=m_Z$ to the GUT scale, $\Lambda_{\rm GUT}=2\times 10^{16}$ GeV. Our results for two loop beta functions are presented in Appendix \ref{betas}. 

For phenomenologically viable points, $y_t$ and $\lambda$ are the largest couplings at the $M_Z$ scale. It is important to notice that the one and two loop contributions to $y_t$ and $\lambda$ in general have numerically opposite signs close to the nonperturbative limit, so it happens that rather than diverging to infinity the couplings reach a fixed point somewhere above 2$\pi$. Given that this fixed point is an artifact of the truncated perturbative series arising close to the non-perturbative limit, we discard viable points for which any of the couplings reaches a value larger than $2\pi$ at $\Lambda_{\rm GUT}$. Because of the cancellation among the 1-loop and 2-loops contributions, $\lambda$ becomes non-perturbative at a value slightly larger than the corresponding value obtained with the one loop beta functions. Among the 13347 viable points collected with the random scan described in the previous section, only 7332, or about half, retain perturbativity at the GUT scale. Among these points, the maximum value of $\left|\lambda\right|$ is 0.85 (0.84 at one loop). Given that  most of the viable perturbative points feature a value of $\left|\lambda\right|$ which is fairly smaller than 0.85, it is important to assess the amount of FT of TESSM at each of these points, and whether this represents an improvement over MSSM.

\begin{figure}[htb]
\centering
\includegraphics[width=0.46\textwidth]{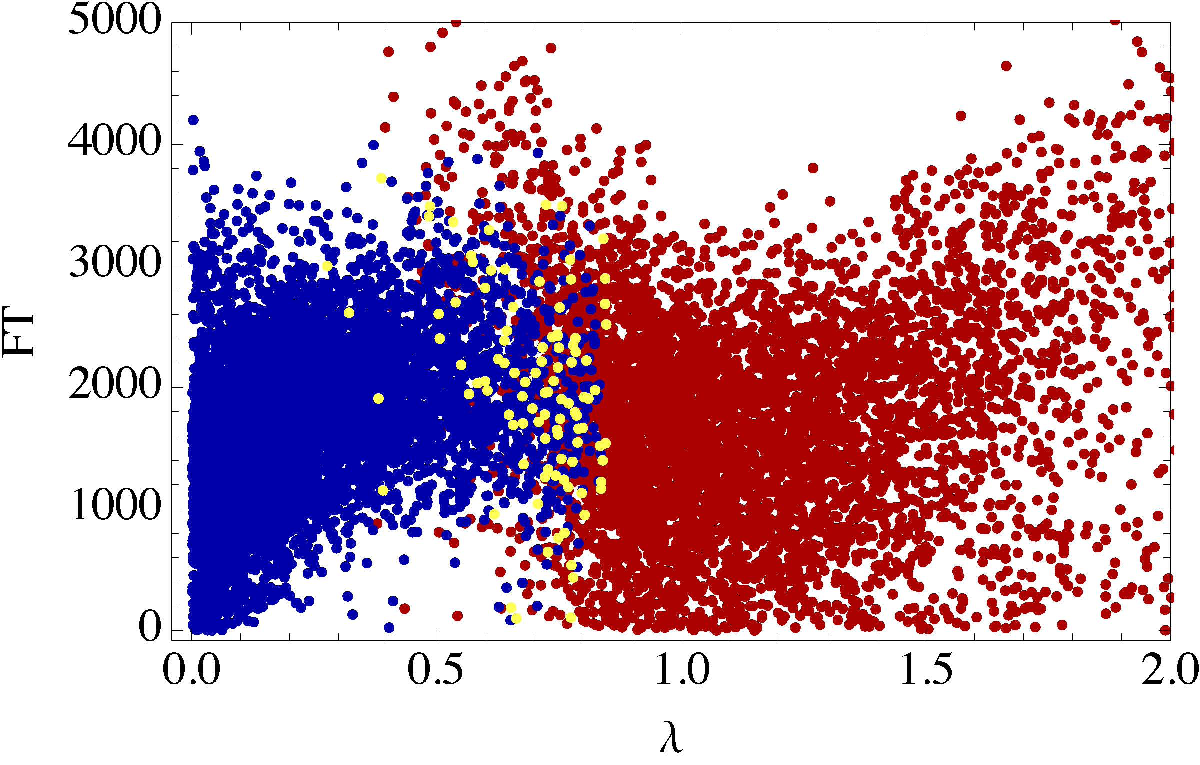}\hspace{0.1cm}
\caption{FT as a function of the triplet coupling $\lambda$: in (red) blue are the (non-perturbative) perturbative points, for which (some) no coupling exceeds $2\pi$ at $\Lambda_{\rm GUT}=2\times 10^{16}$ GeV. In yellow are the points which are perturbative for the two loop but not for the one loop beta functions.}
\label{lambdaFT1}
\end{figure}

A simple estimate of FT in supersymmetry (SUSY) is given by the logarithmic derivative of the EW vev $v_w$ with respect to the logarithm of a given model parameter $\mu_p$ \cite{Ellis:1986yg,Barbieri:1987fn}: this represents the change of $v_w$ for a 100\% change in the given parameter, as defined below:
\be
\text{FT}\equiv \frac{\partial  \log  v_w^2}{\partial  \log  \mu_p ^2 \left(\Lambda\right)  }\ ,\quad\mu_p ^2 \left(\Lambda\right) =\mu_p ^2 \left(M_Z\right)+\frac{\beta _{\mu_p ^2}  }{16 \pi ^2} \log   \left(\frac{\Lambda}{M_Z}\right)\ ,\quad
\beta _{\mu_p ^2}=16 \pi ^2 \frac{d\mu_p ^2}{d\text{logQ}}\ ,
\ee
where in parenthesis is the renormalisation scale of $\mu_p$. In MSSM $v_w$ shows its strongest dependence on $m_{H_u}^2$, which therefore produces also the largest value of FT: this is understandable given that the physical light Higgs is mostly of up type. The value of FT in $m_{H_u}^2$, which we calculate by deriving the one loop beta function of $m_{H_u}^2$, indeed happens to be largest in TESSM as well \footnote{The expression for the FT in $m_{H_d}^2$ becomes non-analytical at $\lambda\sim 0.5$, where there is a pole: excluding the vicinity of this point, for which FT is ill-defined, the largest values of FT indeed are associated to $m_{H_u}^2$.}:
\bea\label{FTdef}
{\rm FT} &=&\frac{\log\left(\Lambda /M_Z\right)}{16\pi \partial_{v_w^2}m_{H_u}^2}\left(6 y_t^2 A_t^2+3 \lambda^2 A_T^2+3 \lambda ^2 m_{H_d}^2+3 \lambda ^2 m_T^2+3 \lambda ^2 m_{H_u}^2-2 g_Y^2  M_1^2-6 g_L^2 M_2^2\right.\\
 &+& 6 m_Q^2 y_t^2 + \left.6 m_{\tilde{t}}^2 y_t^2+6 m_{H_u}^2 y_t^2+ g_Y^2 \left(3 m_{\tilde{b}}^2-m_{H_d}^2-3 m_L^2+3 m_Q^2-6 m_{\tilde{t}}^2+m_{H_u}^2+3 m_{\tilde{\tau}}^2\right)\right) ,\nonumber
\eea
where the derivative in the denominator acts on the expression of $m_{H_u}^2$, Eqs.~\eqref{stabV}.
In Fig.~\ref{lambdaFT1} we present the value of FT evaluated at $\Lambda_{\rm GUT}$, where in blue are the perturbative points, for which no dimensionless coupling exceeds $2\pi$ in absolute value, in yellow are 102 points that are non-perturbative only at one loop, while in red are the nonperturbative points, as determined by the same criterium: it is clear that while values of $\lambda(M_Z)\sim 1$ indeed produce smaller FT, these large values also drive TESSM into a non-perturbative regime. Noticeably, for $\lambda$ values larger than 1 the tree-level mass of the light Higgs easily exceeds 125.5~GeV, in which case a large quantum correction, which drives up FT, is actually necessary to cancel the excess in mass. It is important to point out that when $\lambda\lesssim 0.2$ it is possible to obtain small FT as long as $t_{\beta}$ is large. 

\begin{figure}[htb]
\centering
\includegraphics[width=0.46\textwidth]{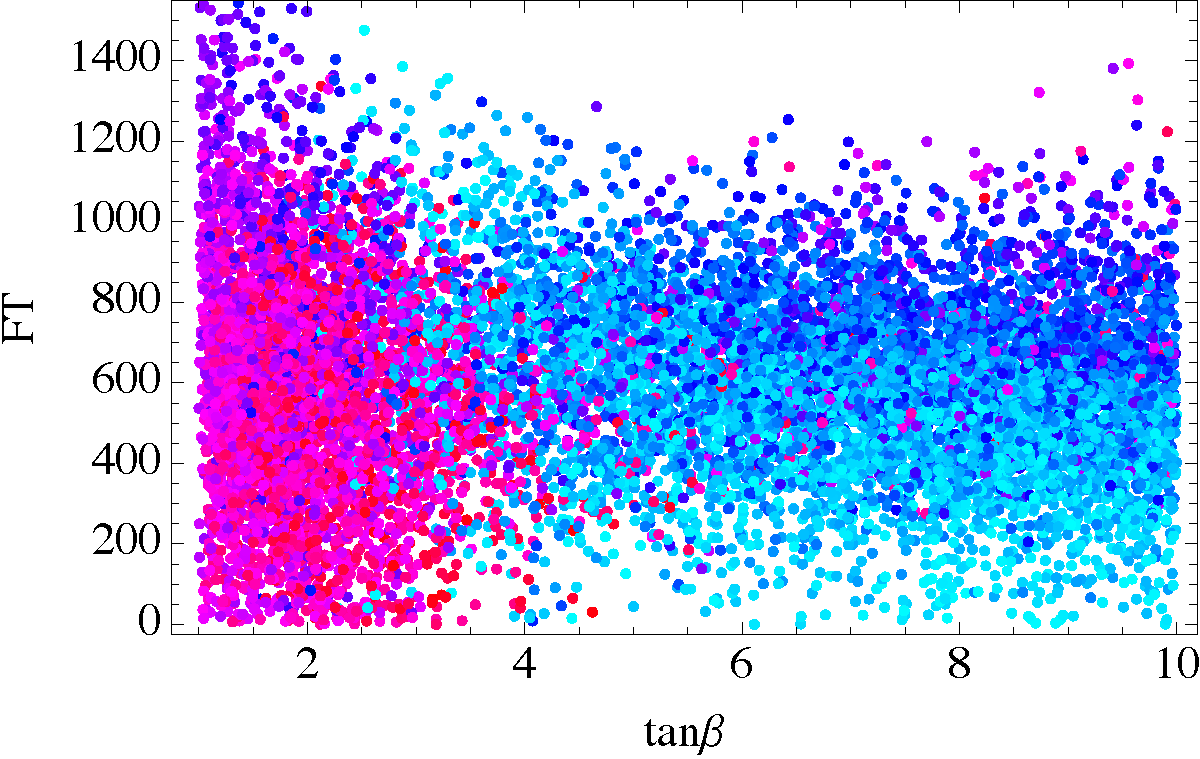}\hspace{1cm}
\includegraphics[width=0.12\textwidth]{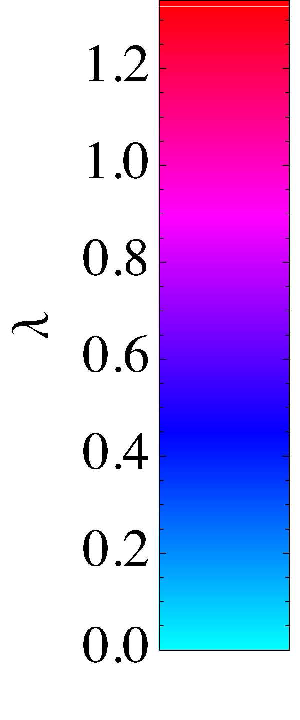}\hspace{0.1cm}
\caption{FT as a function of $\tan\beta$: the region of small $\tan\beta$ and small FT is accessible only for values of $\lambda>0.8$.}
\label{lambdaFT2}
\end{figure}

For $\lambda(M_Z)\sim 1$, the coupling remains perturbative up to scales much higher than the one of $O({\rm TeV})$ tested at LHC. Taking a cutoff scale as high as the GUT scale is indeed less justifiable for TESSM than for MSSM, given that the triplet in the particle content spoils the unification of the gauge couplings at $\Lambda_{\rm GUT}$. Moreover, possible UV completions that generate spontaneous SUSY breaking in TESSM might well also alter the running of $\lambda$. Given these reasons, in the following analysis we choose a less restrictive cutoff scale, $\Lambda_{\rm UV}=10^4$ TeV, which is approximately the highest scale tested experimentally through flavor observables \cite{Beringer:1900zz}. Among the 13347 scanned viable data points, 11244 retain perturbativity at $\Lambda_{\rm UV}$, featuring $|\lambda|\leq 1.34$. In Fig.~\ref{lambdaFT2} we plot the FT associated to each of these viable points in function of $\tan\beta$, with a colour code showing the corresponding value of $|\lambda|$. Values of $\tan\beta$ close to 1 can be reached only for large values of $|\lambda|$ (greater than about 0.8) where the corresponding FT can be considerably smaller than for small values of $|\lambda|$, naively associated to MSSM-like phenomenology. In the same large $|\lambda|$ region, many data points suffer from large FT because  $m_{h^0_1}$ at tree-level is actually much larger than 125.5 GeV, and so a large quantum correction is needed to achieve the right light Higgs mass value. For smaller values of $|\lambda|$ (greater than about 0.5), small $\tan\beta$ solutions also exist in a few cases but they lead to large FT. This is understandable because either $|\lambda|$ is large enough to generate most of the 125.5~GeV light Higgs mass at tree-level, or the stops need to be very heavy to compensate the smallness of $\tan\beta$, which in turn increases FT. 

\begin{figure}[htb]
\includegraphics[width=0.46\textwidth]{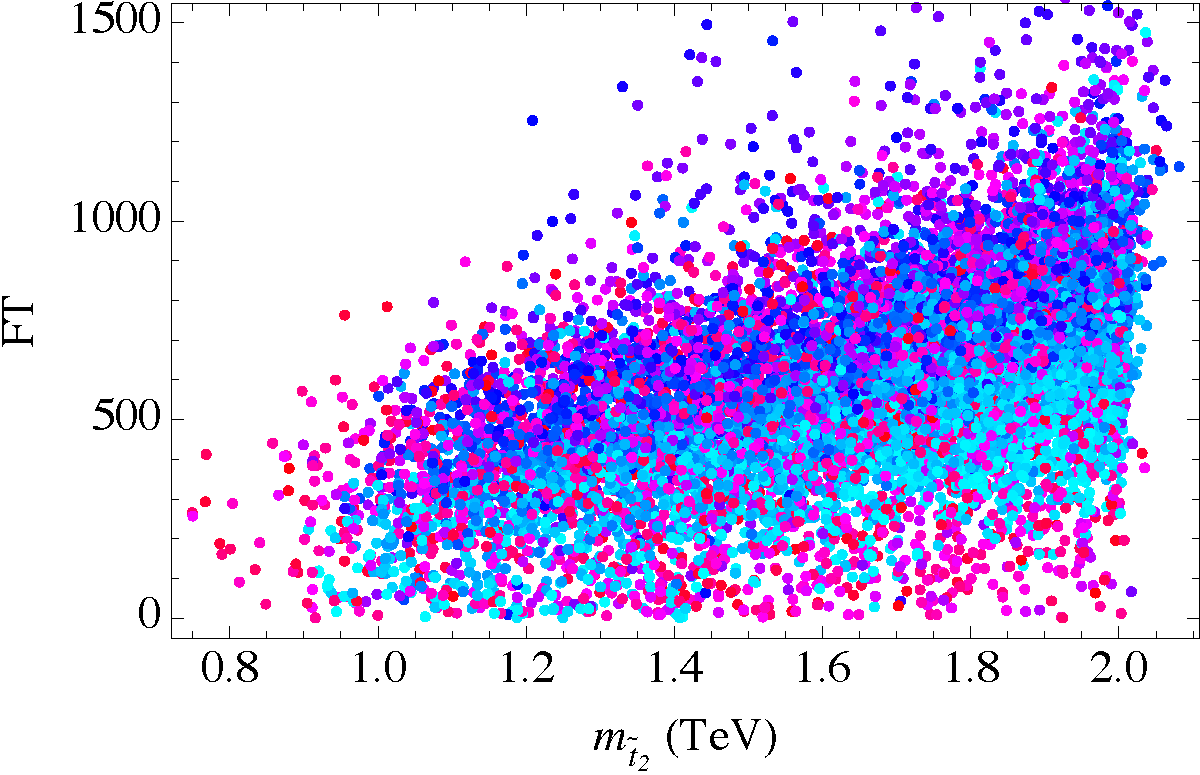}\hspace{1cm}
\includegraphics[width=0.46\textwidth]{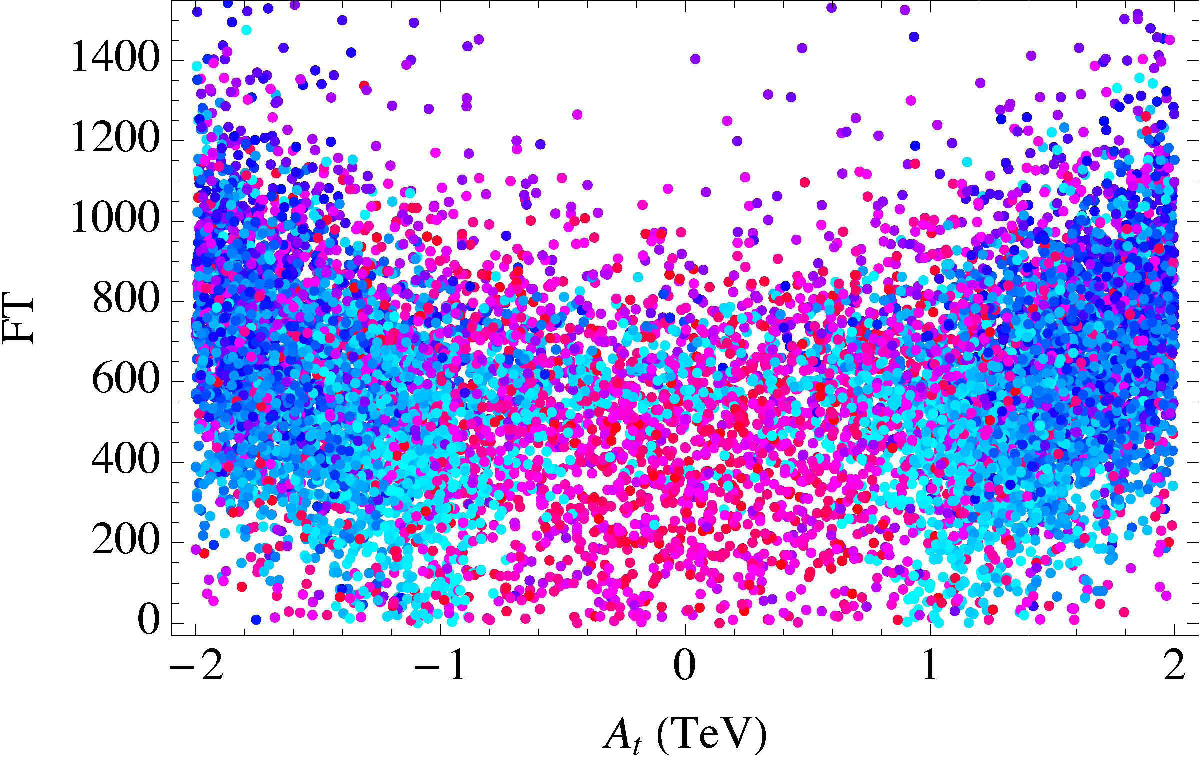}\hspace{0.1cm}
\caption{FT as a function, respectively, of the heavier stop mass $m_{\tilde{t}_2}$ (left panel) and the cubic stop coupling $A_t$ (right panel). Interestingly, for values of $|\lambda|>0.8$ it often happens that the tree-level light Higgs mass exceeds by a large amount 125.5 GeV, in which case another large but negative stop contribution, which generates a large FT, is required for viability. We also notice that the region of small $\left|A_t\right|$ is viable exclusively for values of  $|\lambda|>0.8$, therefore opening up a region unaccessible to MSSM.}
\label{lambdaFT3}
\end{figure}

This pattern is shown in Fig.~\ref{lambdaFT3}, where FT is plotted both as a function of the heavier stop mass and of $A_t$. It is interesting to notice that the viable region of small $\left|A_t\right|$ and small FT, like that of small $\tan\beta$, is accessible only for large values of $|\lambda|$, greater than about 0.8, where $m_{\tilde{t}_2}$ could be large. For small values of $|\lambda|$, $\left|A_t\right|$ needs to be large to generate the measured $m_{h_1^0}$.

In the next Section we define the couplings relevant for light Higgs physics at LHC in terms of a set of coupling coefficients and SM-like couplings, and introduce an additional coupling coefficient of the heavy Higgses necessary to rescale the direct search constraint on the mass of a heavy SM Higgs. Equipped with these tools we then perform a goodness of fit analysis using the current experimental data.

\section{Higgs Physics at LHC}\label{Hphy}
\begin{figure}[htb]
\includegraphics[width=0.46\textwidth]{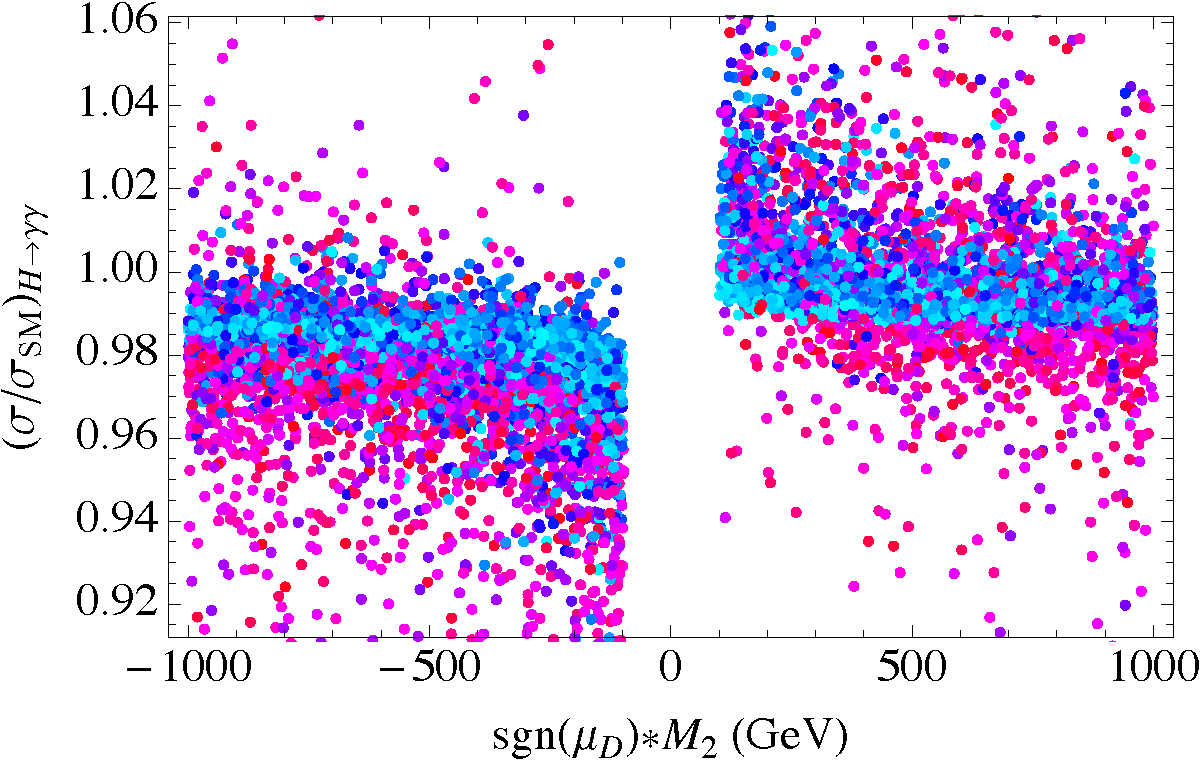}\hspace{1cm}
\includegraphics[width=0.46\textwidth]{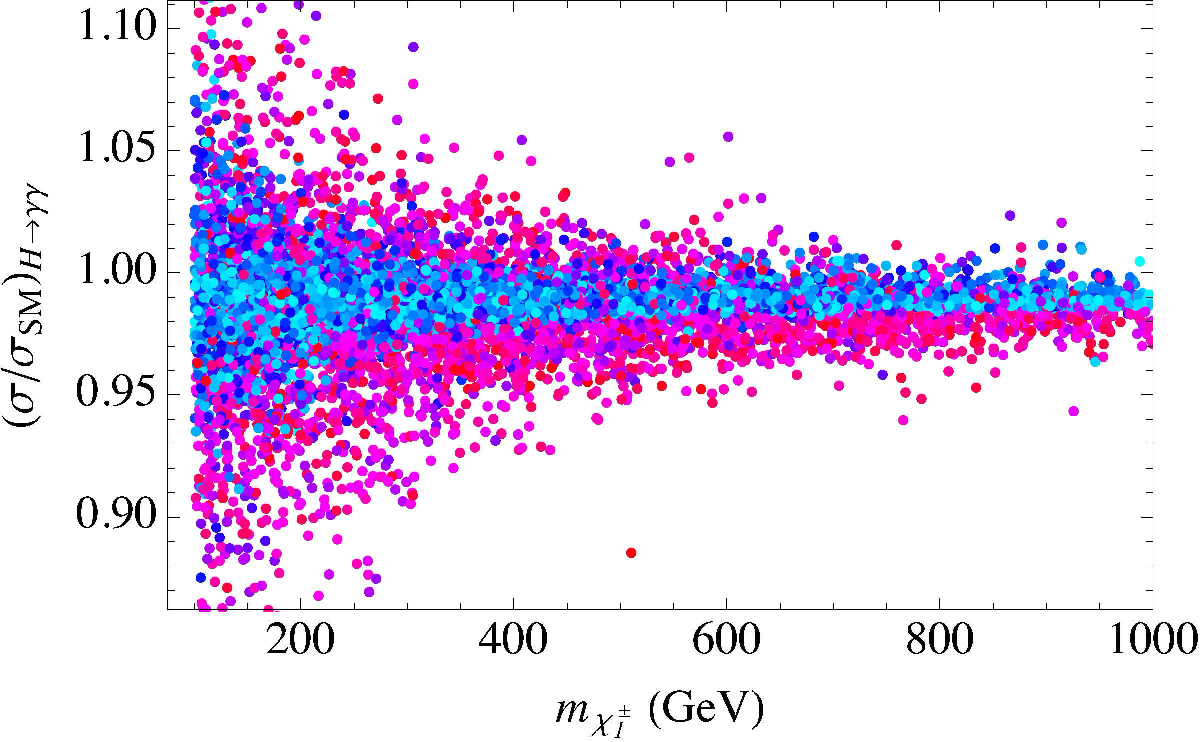}\hspace{0.1cm}
\caption{Higgs decay rate to diphoton of the TESSM relative to the SM as a function, respectively, of ${\rm sign} (\mu_D)\times M_2$ (left panel) and of the lightest chargino mass $m_{\chi_1^\pm}$ (right panel). For opposite signs of $M_2$ and $\mu_D$, most of the viable points feature a suppression of the Higgs decay rate to diphoton as compared to the SM rate. The suppression or enhancement of the decay rate increases with decreasing $m_{\chi_1^\pm}$.}
\label{H2gamma1}
\end{figure}

Among the light Higgs production and decay channels, the only processes for which the non-SM particles become relevant are the gluon-gluon fusion and the decay to $\gamma\gamma$. The total contribution of non-SM particles to these loop-induced processes can be simply accounted for in the effective Lagrangian by adding a coloured and a charged scalar, respectively labeled $\Sigma$ and $S$, with masses much larger than 125.5~GeV. The couplings of these scalars and of the SM particles to the light Higgs can be expressed by rescaling the corresponding SM-like coupling by a coefficient. The light Higgs linear coupling terms that mimic the TESSM contributions to Higgs physics at LHC can therefore be written as\footnote{A similar parametrization of non-SM particles contributions to loop processes has been used in \cite{Antola:2013fka}.}
\bea
{\cal{L}}_{\textrm{eff}} &=& a_W\frac{2m_W^2}{v_w}hW^+_\mu W^{-\mu}+a_Z\frac{m_Z^2}{v_w}hZ_\mu Z^\mu
-\sum_{\psi=t,b,\tau}a_\psi\frac{m_\psi}{v_w}h\bar{\psi}\psi\nonumber \\
&&-a_\Sigma\frac{2m_\Sigma^2}{v_w}h\Sigma^* \Sigma-a_S\frac{2m_S^2}{v_w}hS^+ S^-.
\label{efflagr}
\eea
The experimental results are expressed in terms of the signal strengths, defined as
\be
\hat{\mu}_{ij}=\frac{\sigma_{\textrm{tot}}{\textrm{Br}}_{ij}}{\sigma_{\textrm{tot}}^{\textrm{SM}} \textrm{Br} ^{\textrm{SM}}_{ij}}\ ,\quad \sigma_{\rm tot}=\sum_{\Omega=h,qqh,\ldots}\!\epsilon_\Omega\sigma_{\Omega} \ ,
\label{LHCb}\ee
where $\textrm{Br}_{ij}$ is the light Higgs branching ratio into the $ij$ particles, $\sigma_\Omega$ the production cross section of the given final state $\Omega$, and $\epsilon_\Omega$ is the corresponding efficiency, which for inclusive searches is equal to 1. The production cross sections and decay rates for tree-level processes in TESSM are straightforwardly derived from Eqs.~(\ref{efflagr}, \ref{LHCb}) by rescaling the corresponding SM result with the squared coupling coefficient of the final particles being produced. For loop induced processes the calculation is more involved. By using the formulas given in \cite{Gunion:1989we} we can write\footnote{In the Eqs.~(\ref{hgamgam}, \ref{hgluglu}) we drop all the labels of $h$ given that these formulas apply generically to any SM-like Higgs particle.}
\be\label{hgamgam}
\Gamma_{h\rightarrow \gamma\gamma}= \frac{\alpha_e^2 m_{h}^3}{256 \pi^3 v_w^2}\left| \sum_i  N_i e^2_i a_i F_{i} \right|^2,
\ee
where the index $i$ is summed over the SM charged particles plus $S^\pm$, $N_i$ is the number of colours, $e_i$ the electric charge in units of the electron charge, and the factors $F_{i}$ are defined by
\bea\label{dpam}
F_{W}&=&\left[2+3 \tau_{W}+3 \tau_{W}\left( 2-\tau_{W} \right) f(\tau_{W})\right]\,;\nonumber\\
F_{\psi}&=&-2 \tau_{\psi}\left[1+\left( 1-\tau_{\psi} \right) f(\tau_{\psi})\right]\ ,\quad \psi=t,b,\tau,c\ ;\nonumber\\
F_{S}&=&\tau_{S}\left[ 1-\tau_{S}  f(\tau_{S})\right] ,\ \, \tau_{i}=\frac{4 m_i^2}{m_{h}^2}\,,
\eea
with
\bea
f(\tau_{i})=\left\{
\begin{array}{ll}  \displaystyle
\arcsin^2\sqrt{1/\tau_{i}} & \tau_{i}\geq 1 \\
\displaystyle -\frac{1}{4}\left[ \log\frac{1+\sqrt{1-\tau_{i}}}
{1-\sqrt{1-\tau_{i}}}-i\pi \right]^2 \hspace{0.5cm} & \tau_{i}<1
\end{array} \right. .
\label{eq:ftau}
\eea
In the limit of heavy  $S^\pm$, one finds
\be
F_S=-\frac{1}{3}\ .
\label{fwps}\ee
We account for the contribution to Higgs decays to diphoton of the charged non-SM particles in TESSM by defining
\be\label{aSd}
a_S \equiv -3 \left[ \sum^3_i \left(F_{h_i^\pm}+F_{\chi_i^\pm}\right)+\sum^2_j \left(\frac{4}{3} F_{\tilde{t}_j}+\frac{1}{3} F_{\tilde{b}_j}\right)\right]\ ,
\ee
where the functions $F$ for scalars and fermions are given by Eqs.~\eqref{dpam} after proper relabelling. Similarly to the two photon decay, the light Higgs decay rate to two gluons is given by
\be\label{hgluglu}
\Gamma_{h\rightarrow g g}= \frac{\alpha_s^2 m_{h}^3}{128 \pi^3 v_w^2}\left| \sum_i  a_iF_{i} \right|^2\ ,\quad i=t,b,c,\Sigma\ ,
\ee
where the functions $F$ are given by Eqs.~\eqref{dpam} with proper relabelling. An overall factor accounting for the next to leading order QCD contributions \cite{Djouadi:2005gi} is independent of the coupling coefficients in Eq.~\eqref{efflagr}, and so it cancels out in the corresponding ratio of branching ratios in Eq.~\eqref{LHCb}. Similarly to the coupling coefficient $a_S$, to account for the contribution of non-SM particles of TESSM to the light Higgs decay into two gluons, we define $a_\Sigma$ as
\be\label{aSigmad}
a_\Sigma \equiv -3 \sum^2_{j=1} \left(F_{\tilde{t}_j}+F_{\tilde{b}_j}\right)\ .
\ee
To rescale the lower limit on the mass of the heavy neutral Higgs we calculate also $a^\prime_g$, the ratio of the TESSM decay rate, of $h^0_2$ to a gluon pair, to that of a SM-like Higgs of mass $m_{h^0_2}$,
\be
a^\prime_g\equiv\frac{\Gamma_{h^0_2\rightarrow g g}}{\Gamma^{SM}_{h\rightarrow g g}}\ .
\ee
This is still determined by Eqs.~(\ref{hgluglu}, \ref{aSigmad}), evaluated for the coupling coefficients and mass of $h^0_2$, rather than $h_1^0$, and then divided by the corresponding SM result. The most stringent limit on the mass of a heavy SM-like Higgs,  $m_{h^0}>770$ GeV, comes from the gluon-gluon fusion Higgs production, subsequently decaying to $ZZ$ \cite{CMS:2013ada}. Assuming $h^0_2$ to decay on-shell and to be much heavier than twice the $W$ mass, the production rate by gluon-gluon fusion scales like the inverse of the Higgs squared mass, with a branching ratio to vector bosons greater than 0.8 for a SM-like Higgs \cite{Djouadi:2005gi}. Making the further assumption, for simplicity, that the same branching ratio for $h_2^0$ is unitary, which makes the constraint clearly more stringent, we impose
\be\label{Hconst}
a^\prime_g \frac{\left(770\ {\rm GeV}\right)^2}{m^2_{h^0_2}}<0.8\ .\quad 
\ee
We evaluate Eq.~\eqref{Hconst} for each viable data point, and find it to hold for 10957 out of the 11244 viable data points that already satisfy perturbativity constraints. At each of these remaining viable points we then evaluate Eq.~\eqref{hgamgam}, making sure that the fermion mass parameter of each mass eigenstate appears with a negative sign in the Lagrangian, given that this is the convention we use in deriving Eq.~\eqref{hgamgam} \cite{Gunion:1989we}, and if that is not the case, we apply a phase rotation to the corresponding fermionic mass eigenstate to flip the sign of its mass operator. In Fig.~\ref{H2gamma1} we show the value of the Higgs decay rate to diphoton for TESSM  relative to the SM one, as a function of $ \text{sign}\left(\mu_D\right)\times M_2$, the soft wino mass parameter times the sign of the superpotential doublet mass parameter. The colour code, given in Fig.~\ref{lambdaFT2}, shows the $\left|\lambda\right|$ value corresponding to the plotted data point. A possible experimental evidence for a suppression or enhancement of the SM Higgs decay rate to diphoton would point decisively, within TESSM, to an opposite or same sign of $M_2$ relative to $\mu_D$, respectively, besides likely large values of $\lambda$, depending on how large the deviation from the SM prediction is. These two mass parameters contribute to the lightest chargino mass, on which the Higgs decay rate to diphoton is strongly dependent. 

As Fig.~\ref{H2gamma1} (right panel) shows, a small mass for the lightest chargino produces a sizable contribution to the decay rate to two photons, as expected, but this contribution can be either constructive or destructive with the one from the $W$ boson: the latter result seems to be in disagreement with results appeared in previous works on the same triplet extension of MSSM that we study here \cite{DiChiara:2008rg,Delgado:2012sm,Delgado:2013zfa,Arina:2014xya}. 

\begin{figure}[htb]
\centering
\includegraphics[width=0.46\textwidth]{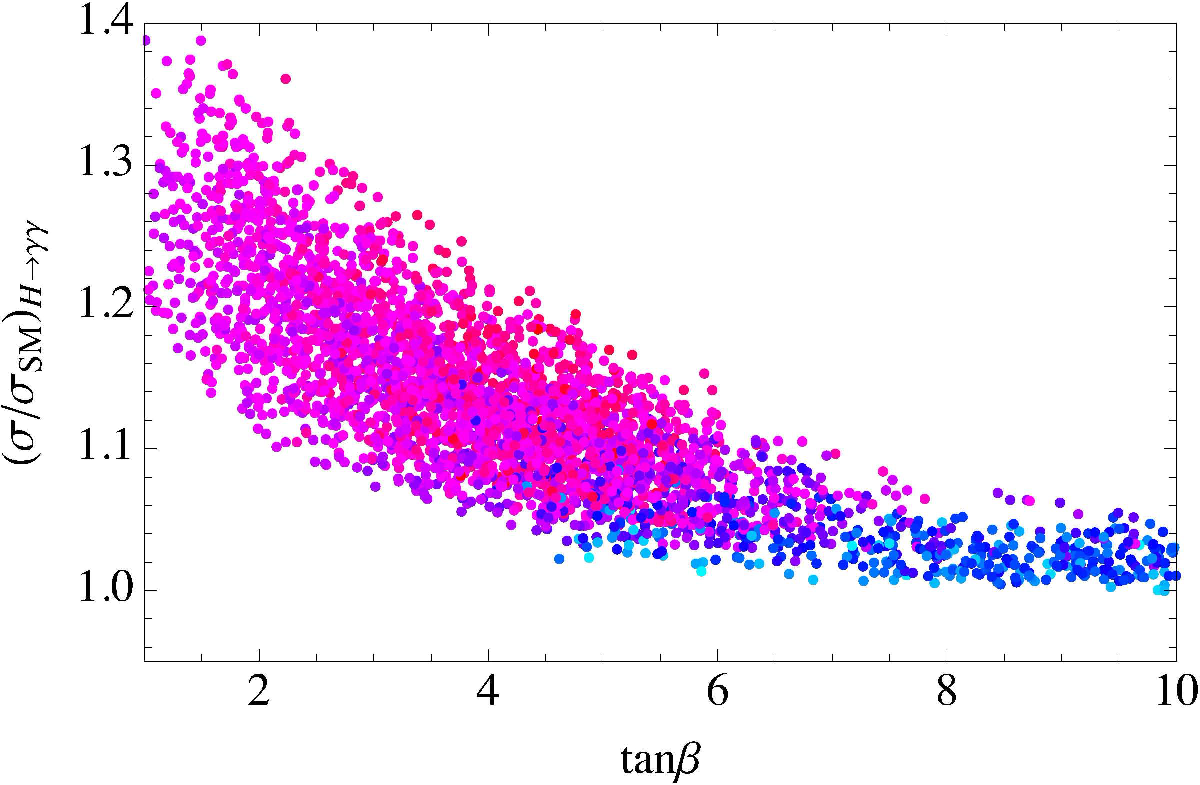}\hspace{0.1cm}
\caption{Higgs decay rate to diphoton in the TESSM relative to the SM as a function of $\tan\beta$ for viable data points scanned only in the positive region of the mass parameters and of the couplings, with a generally small light chargino mass: in this region only an enhancement of the SM decay rate is observed.}
\label{H2gammachk}
\end{figure}

It turns out that the constructive interference is a result of the choice to scan only a specific region of parameter space (positive fermion mass parameters and $\lambda$ coupling) for which the mass term of the mostly triplino-like chargino and the coupling to the light Higgs, unlike the top quark, have opposite signs. As a way of comparison with \cite{DiChiara:2008rg,Delgado:2012sm,Delgado:2013zfa,Arina:2014xya} we scan the parameter region again for viable points within the region defined below 
\bea\label{ps}
&&1\leq t_{\beta }\leq 10\ ,\ 0\leq \lambda \leq 1\ ,\ 5 \,\text{GeV}\leq \mu _D,\mu _T\leq 250\,\text{GeV}\ ,\ 50 \,\text{GeV}\leq M_1,M_2\leq 300 \,\text{GeV}\ ,\nonumber\\ 
&& A_t=A_T=B_T=0\, ,\, 0\,\leq B_D\leq 2 \,\text{TeV}\ ,\ 500 \,\text{GeV}\leq m_Q,m_{\tilde{t}},m_{\tilde{b}}\leq 2 \,\text{TeV}
\eea
which roughly corresponds to (and exceeds) the region scanned in \cite{Arina:2014xya}, and apply again the perturbativity constraints (no coupling larger than $2\pi$ at $\Lambda_{\rm UV}$) as well as the lower bound on $m_{h^0_2}$, Eq.~\eqref{Hconst}. One key difference with previous calculations is that, among the non-SM particles, we include in the decay rate to diphoton the contributions of all the third generation SM and non-SM charged particles, without making any assumption on the coupling coefficients or masses of these particles. The result of the scan of this region of the TESSM parameter space is shown in Fig.~\ref{H2gammachk}. It is clearly consistent with previous results, as it shows that in this region of the parameter space only an enhancement, which becomes comparably large with large positive values of $\lambda$, is possible. 

In the next Section we calculate a low energy flavor observable, $\mathcal{B}r(B_s\to X_s \gamma)$, which provides a strong constraint on the absolute size of $\lambda$.

\section{$\mathcal{B}r(B_s\to X_s \gamma)$ in TESSM}\label{btsgsec}

Besides the constraints obtained from Higgs decay channels, the low energy observables also provide stringent limitations on the parameter space of new physics beyond the SM. In particular, the parameter space of MSSM-like models with minimal or general flavour mixings in the sfermion sector has been investigated in great detail with the help of $B$-physics observables \cite{Bphysics}. Recently, it has been pointed out in Ref. \cite{Btomumu} that the branching ratio of the flavour changing decay $B_s\rightarrow X_s \gamma$ plays a very important role in constraining the viable parameter space of MSSM especially for low $\tan\beta$, whereas the flavour bounds obtained from the branching ratio of $B_s\rightarrow \mu^+ \mu^-$ become relevant only for large values of $\tan\beta$ ($\gtrsim 10$). Since we limit our phenomenological study of TESSM to the low $\tan\beta$ region, it is sufficient to consider only the $B_s\rightarrow X_s \gamma$ decay for the rest of the analysis.

For any model, the branching ratio of  $B_s\to X_s\gamma$ can be calculated via the effective Hamiltonian approach described by the generic structure   

\be
\mathcal{H}_{eff}= \frac{G_F}{\sqrt{2}}V^*_{ts}V_{tb}\sum_i C_{i}(\mu_r)Q_{i}(\mu_r)\ ,
\label{effham}
\ee
where $V_{ij}$ are the entries of the CKM matrix, $C_{i}$ the Wilson coefficients, $\mu_r$ the renormalization scale, and $Q_i$ the relevant dimension 6 local operators. Here the Wilson coefficients can be written in the following form

\bea\label{wilson}
C_i (\mu_r)&=& C_i^{(0){\rm SM} } (\mu_r)+C_i^{(0)h_i^{\pm}} (\mu_r)+C_i^{(0){\rm SUSY}} (\mu_r)\nonumber\\
&+&\frac{\alpha_s(\mu_r)}{4 \pi}\left( C_i^{(1){\rm SM}} (\mu_r)+C_i^{(1)h_i^{\pm}} (\mu_r)+C_i^{(1){\rm SUSY}} (\mu_r)\right) \nonumber.
\eea
where $C_i^{(0)}$ stands for the leading order corrections (LO) to the Wilson coefficients while $C_i^{(1)}$ represents the next to leading order (NLO) effects. In particular, for $C_{i}^{(0){\rm SUSY}}$ we only consider the corrections from 1-loop chargino diagrams, in $C_i^{(1){\rm SUSY}}$ we include the 2-loops contributions of the three charginos and the gluino \cite{NLOSUSYbsg}, while those of the three charged Higgses are given by $C_i^{(1)h_i^{\pm}}$. 

Similarly, the leading and next to leading order contributions from the SM at the $M_W$ scale can be obtained from \cite{LOSMCH}. For the charged Higgs contributions, Ref. \cite{LOSMCH} can be used as a starting point where one needs to replace the charged Higgs-quark couplings of the MSSM with the ones in TESSM: given that the latter possesses three physical charged Higgses, their contributions are summed over. After the total contribution at the $M_W$ scale is obtained, Ref.\cite{wilsonrun} can be used as a guideline to calculate the Wilson coefficients at the desired scale $\mu_r$. Here we emphasize that even though there is a greater number of particles that contribute to  $B_s\rightarrow X_s \gamma$, it is still possible to get some suppression in the corresponding branching ratio, compared with the MSSM one, because of the lack of triplet coupling to the SM fermions. 
In other words, the physical charged Higgses and charginos with triplet components give a suppressed contribution, as compared to their MSSM counterparts,  to the rare $B$ decays to $ X_s\gamma$.

\begin{figure}
\centering
\includegraphics[width=0.46\textwidth]{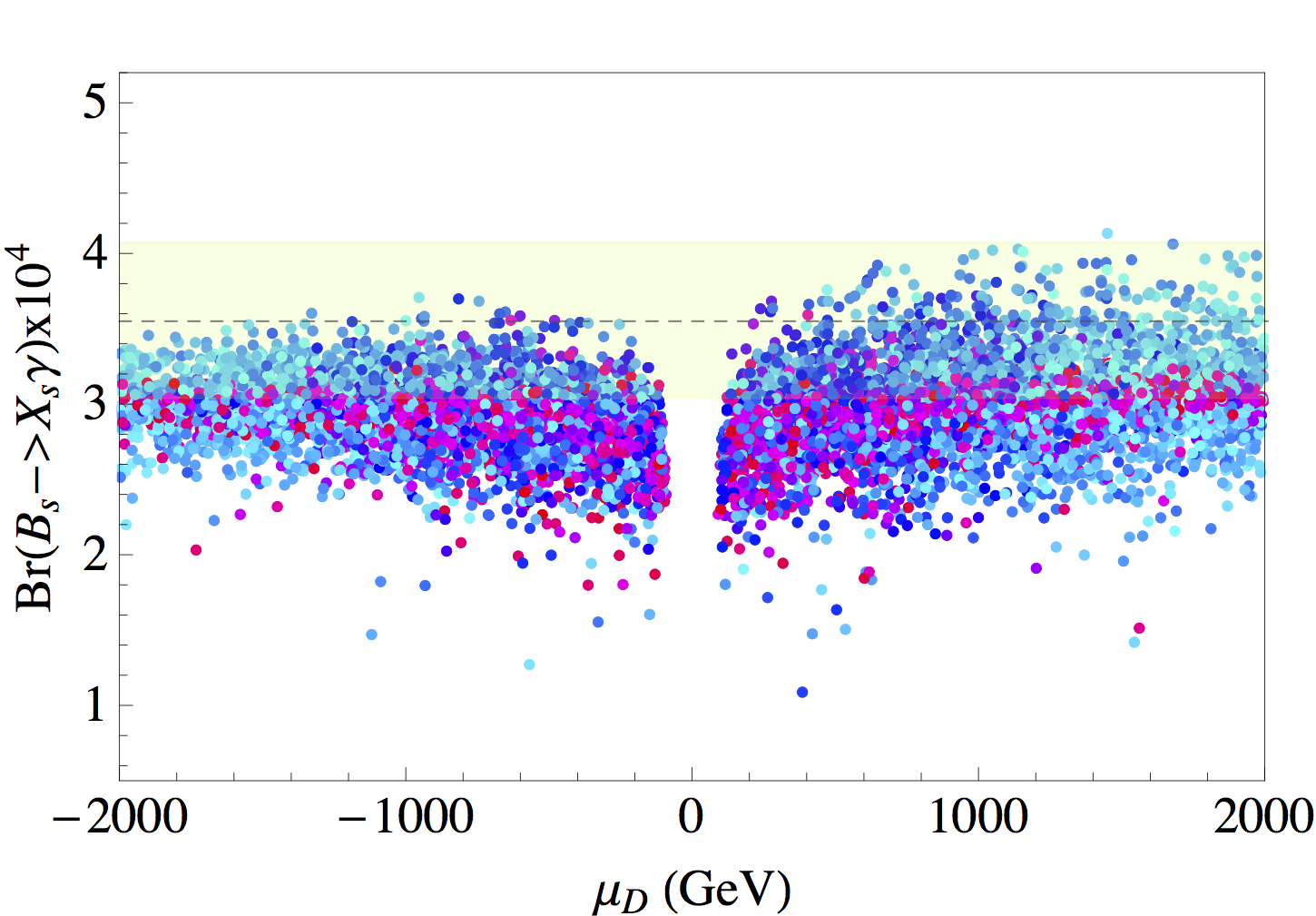} 
\caption{Values of $\mathcal{B}r(B_s\to X_s\gamma$) associated to each viable data point as a function of $\mu_D$, where the NLO SUSY effects are taken into account. The yellow band shows the viable region at the $2\sigma$ CL around the experimental value of $\mathcal{B}r(B_s\to X_s\gamma)$.}\label{bsgmu}
\end{figure}

For the numerical analysis we calculate, at the next to leading order (NLO) and within TESSM, the values of $\mathcal{B}r(B_s\to X_s\gamma)$ corresponding to each of the 10957 viable data points, featuring perturbativity up to $\Lambda_{\rm UV}=10^4$ TeV, defined in Sections~\ref{finetuningTESSM}, \ref{Hphy}.
In Fig.~\ref{bsgmu} we plot $\mathcal{B}r(B_s\to X_s\gamma)$ as a function of $\mu_D$, and we use the colour code defined in Fig.~\ref{lambdaFT2} to represent different values of $\lambda$.
For small $|\mu_D|$, the contribution coming from the chargino with a mass mostly proportional to $\mu_D$ is non-negligible and, depending on the sign of $A_t$, this contribution increases or diminishes the total contribution to the $B_s\to X_s\gamma$ branching ratio.  We observe that for values of $|\mu_D|\gsim$ 1 TeV a majority of data points fall within $\pm 2 \sigma$ of the experimental value, with $\mathcal{B}r(B_s\to X_s\gamma)_{exp}=3.55\pm0.24\pm0.09\times 10^{-4}$, and with $|\lambda|$ being generally small. It is relevant to point out that at LO $\mathcal{B}r(B_s\to X_s\gamma)$ is symmetric with respect to the sign of $\mu_D$, while at NLO there is a clear preference for the positive sign of $\mu_D$.


In order to understand this low $\lambda$ preference we investigate the effect of the mass as well as the structure of the lightest charged Higgs on the $\mathcal{B}r(B_s\to X_s\gamma)$. A large majority (93\%) of the viable data points with small lambda ($|\lambda|\leq 0.6$) features a larger triplet than doublet component of the lightest chargino mass eigenstate. This in turn produces a suppression of the $B_s\rightarrow X_s \gamma$ branching ratio, given that the triplet field gives no contribution at NLO to the $B_s\rightarrow X_s \gamma$ decay because it lacks direct couplings to quarks. For large $\lambda$ values, the $B_s\to X_s\gamma$ branching ratio falls within $2\sigma$ of the experimental value only for $m_{h_1^{\pm}}\gsim 700$ GeV, since the negative contribution of $h^\pm_1$ to the branching ratio becomes smaller in absolute value as $m_{h_1^{\pm}}$ increases.

\begin{figure}
\centering
\includegraphics[width=0.46\textwidth]{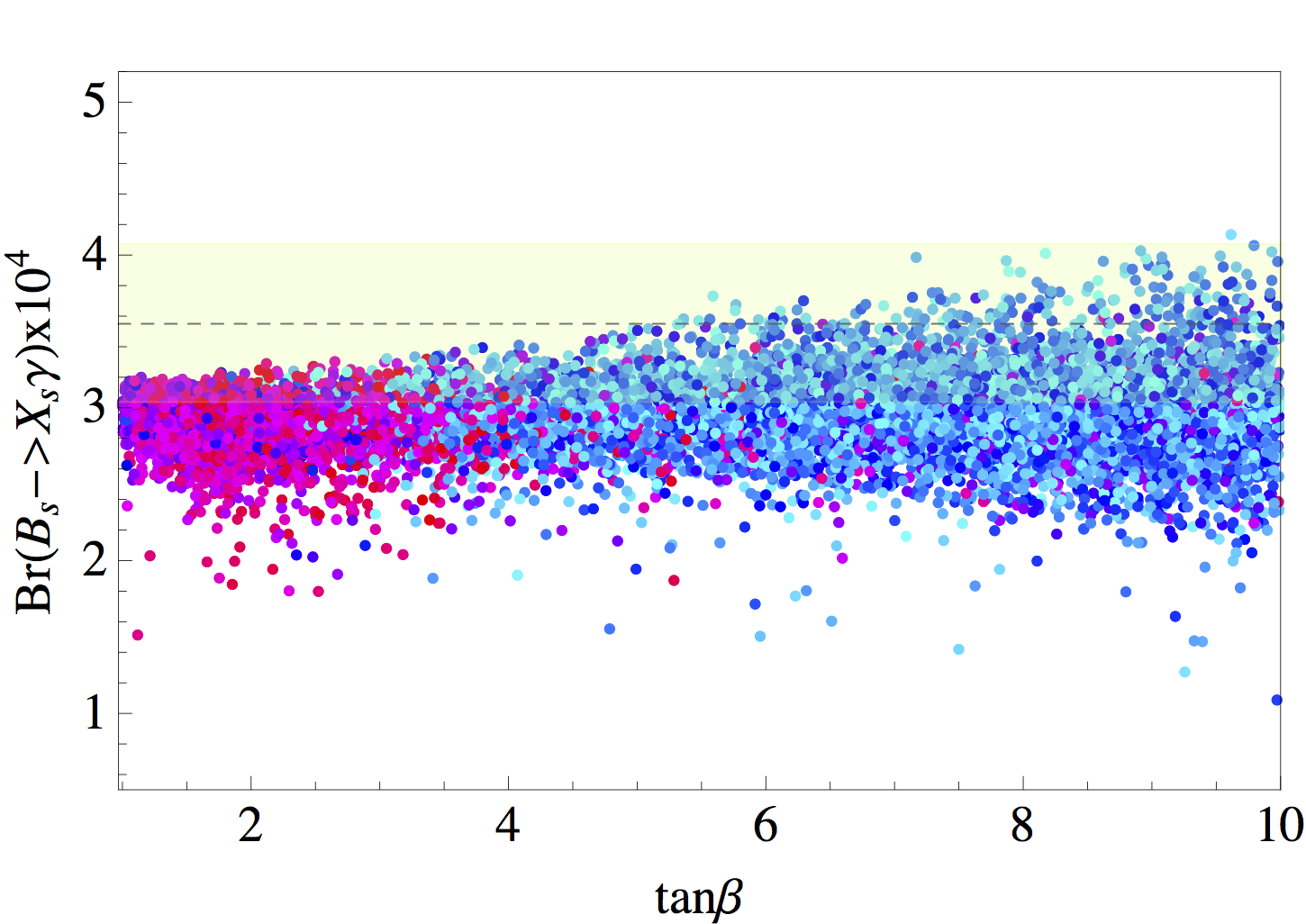}
\caption{The values of $\mathcal{B}r(B_s\to X_s\gamma$) for the allowed data points as a function of $\tan\beta$. The yellow band represents the viable region at $2\sigma$ CL around the experimental value of $\mathcal{B}r(B_s\to X_s\gamma)$.}\label{bsgtan}
\end{figure}

Next we illustrate the $\tan \beta$ dependence of $\mathcal{B}r(B_s\to X_s\gamma)$, plotted in Fig.~\ref{bsgtan}. For values of $\tan\beta$ close to 10, corresponding to small values of $\lambda$, about half of the data points feature a $\mathcal{B}r(B_s\to X_s\gamma)$ prediction within $\pm2\sigma$ of the experimental value, while the other half generates a suppressed branching ratio. 
For low $\tan\beta$ values, corresponding to large $\lambda$, the $\mathcal{B}r(B_s\to X_s\gamma)$ values associated to the viable data points sit mostly below the lower $2\sigma$ bound, and for no point the prediction actually matches the experimental value. It seems that the very large $\lambda$ values favored by FT, as discussed in Section~\ref{finetuningTESSM}, are severely constrained by the $B_s\to X_s\gamma$ branching ratio. This clear preference of the experiment for smaller values of $|\lambda|$ is unwelcome, given that, as shown in Section~\ref{finetuningTESSM}, values of $|\lambda|$ close to 1 can greatly reduce the amount of FT. On the other hand there are other observables, like the Higgs decay rate to diphoton, which prefer large values of $|\lambda|$, and can therefore tip the balance in favor of low FT. In the next Section we perform a goodness of fit analysis on the Higgs physics observables detailed in Section~\ref{Hphy} as well as $\mathcal{B}r(B_s\to X_s\gamma)$.

\section{Goodness of Fit to LHC Data}
\label{fitsect}

To determine the experimentally favored values of the free parameters $a_W,a_Z,a_u,a_d,a_S,a_\Sigma$, we minimize the quantity
\be
\chi^2=\sum_i\left(\frac{{\cal O}_i^{\textrm{exp}}-{\cal O}_i^{\textrm{th}}}{\sigma_i^{\textrm{exp}}}\right)^2,
\label{chi2}\ee
where $\sigma_i^{\textrm{exp}}$ represent the experimental uncertainty, while the observables ${\cal O}_i^{\textrm{exp}}$ correspond to the signal strengths, defined by Eq.~\eqref{LHCb}, for Higgs decays to $ZZ$, $W^+W^-$, $\tau^+\tau^-$, $b\bar{b}$, as well as all the topologies of decays to $\gamma\gamma$, respectively measured by ATLAS \cite{ATLAS:2013nma,ATLAS:2013wla,ATLAS:2012dsy,ATLAS:2012aha,ATLAS:2013oma} and CMS \cite{CMS:xwa,CMS:bxa,CMS:utj,CMS:2014ega}, and by Tevatron for decays to $W^+W^-$ and $b\bar{b}$ \cite{Aaltonen:2013kxa}. Because of the smallness of the triplet vev, $v_T$, and the relatively large mass of the lightest neutral Higgs, the values of $a_Z$ and $a_W$ for the viable data points are very close to one ($\sim 0.997$). We therefore set $a_W=a_Z=1$ in the $\chi^2$ function defined in \eqref{chi2}. Moreover, given that $a_u$ and $a_d$ are correlated through $\tan\beta$, in the minimization of $\chi^2$ with free coupling coefficients we also set $a_u=a_d=a_f$. The free coupling coefficients $a_f$, $a_S$, and $a_\Sigma$ produce a minimum of $\chi^2$ defined by
\bea\label{optimusprime}
\chi^2_{min}/d.o.f.&=&0.98\, ,\ d.o.f.=55\,,\ p\left(\chi^2>\chi^2_{min}\right)=51\%\, ,\nonumber\\
 \hat{a}_f&=&1.03\,,\ \hat{a}_S=-2.30\,,\ \hat{a}_\Sigma=-0.04\ .
\eea
As a way of comparison, we determine the corresponding results for the SM, which has no free parameters:
\be
\chi^2_{min}/d.o.f.=0.96\, ,\ d.o.f.=58\,,\ p\left(\chi^2>\chi^2_{min}\right)=56\%\, .
\ee
One can define an approximate expression of $\chi^2$ around its minimum by assuming that the deviations of the free coupling coefficients from their optimal values (denoted by a hat in Eq.~\eqref{optimusprime} and below) are small as compared with their respective uncertainties \cite{Giardino:2013bma}:
\be
\Delta \chi^2=\chi^2-\chi^2_{min}=\delta^T \rho^{-1} \delta\,,\ \delta^T=\left( \frac{a_f-\hat{a}_f}{\sigma_f}, \frac{a_S-\hat{a}_S}{\sigma_S}, \frac{a_\Sigma-\hat{a}_\Sigma}{\sigma_\Sigma} \right)\ ,
\ee
with
\be
\sigma_f=0.165\,,\ \sigma_S=2.79\,,\ \sigma_\Sigma=0.431\,,\ \rho=
\left(\begin{array}{ccc}1 & -0.6 &  -0.685 \\-0.6 & 1 & 0.785 \\ -0.685 & 0.785 & 1\end{array}\right)\ ,
\ee
where the uncertainties are explicitly defined to correspond to $\Delta\chi^2=1$. 

\begin{figure}[htb]
\includegraphics[width=0.46\textwidth]{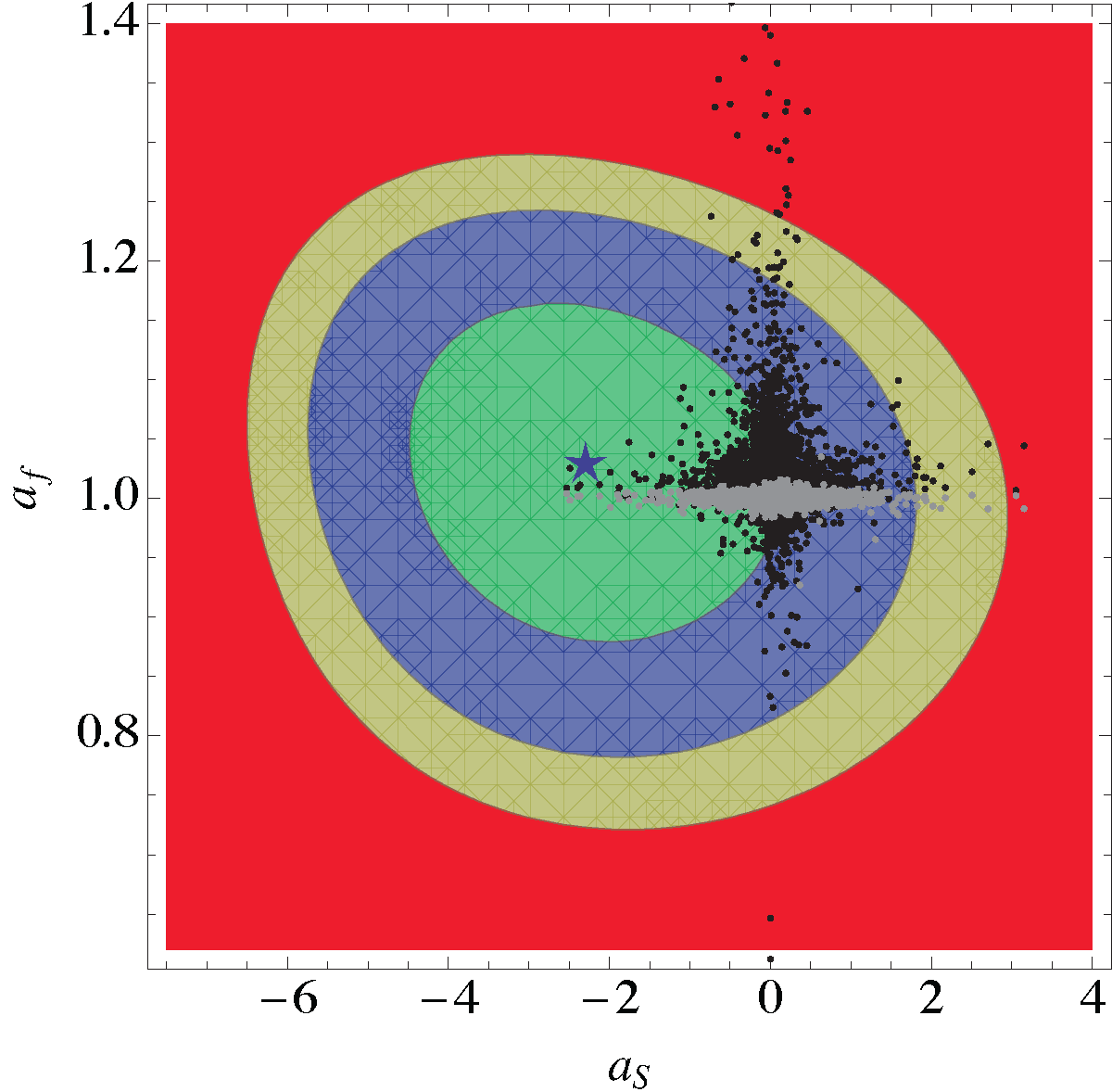}\hspace{1cm}
\includegraphics[width=0.46\textwidth]{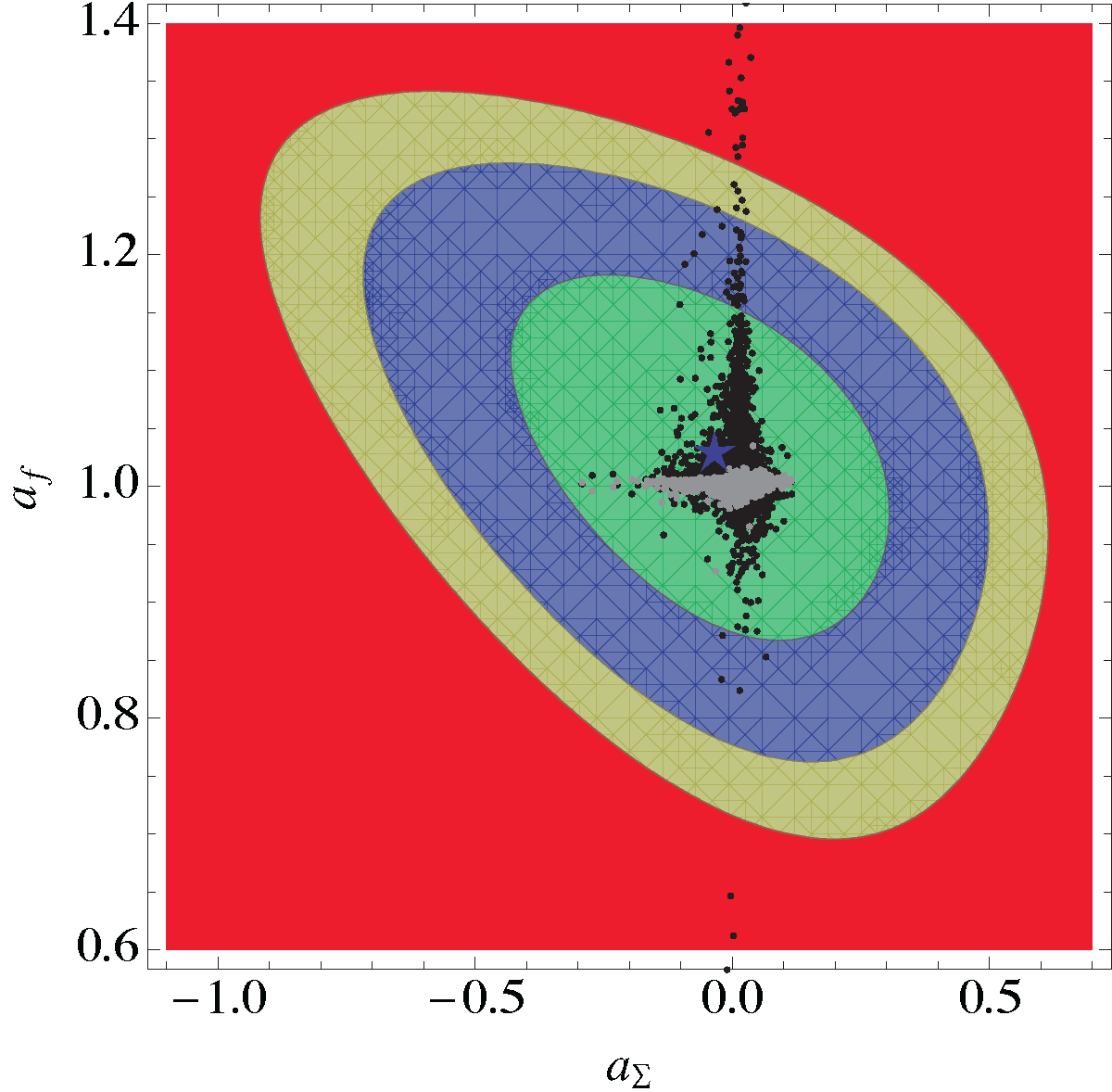}\hspace{0.1cm}
\caption{Viable regions at the 68\%, 95\%, 99\% CL in the coupling coefficients $a_S,a_f$ (left panel) and $a_\Sigma,a_f$ (right panel) planes passing through the optimal point (blue star), together with the values of $a_u$ (grey) and $a_d$ (black) associated with each viable point.}
\label{aSafaSigmaaf}
\end{figure}

In calculating $\chi^2$ for the TESSM viable data points we include also the $\mathcal{B}r(B_s\to X_s \gamma)$ observable. Assuming a total of four free parameters ($a_f,a_S,a_\Sigma$, plus one more to fit $\mathcal{B}r(B_s\to X_s \gamma)$), the viable data point featuring minimum $\chi^2$ has
\be
\chi^2_{min}/d.o.f.=1.01\, ,\ d.o.f.=55\,,\ p\left(\chi^2>\chi^2_{min}\right)=46\%\ .
\ee 
This result should be compared with the SM one for the same set of observables:
\be
\chi^2_{min}/d.o.f.=0.99\, ,\ d.o.f.=59\,,\ p\left(\chi^2>\chi^2_{min}\right)=50\%\, .
\ee
We notice that the goodness of fit of TESSM is comparable, although smaller, to that of the SM. It is important, however, to realize that the quoted $p$ values are only indicative of the viability of TESSM and SM relative to one another, given that the chosen set of observables, besides $\mathcal{B}r(B_s\to X_s \gamma)$, tests only the linear Higgs sector of the Lagrangian. In Figs.~\ref{aSafaSigmaaf} we plot the 68\%, 95\%, 99\% CL viable regions (respectively in green, blue, and yellow) on the planes $a_S-a_f$ and $a_\Sigma-a_f$, each intersecting the optimal point (blue star) defined in Eq.~\eqref{optimusprime}. On the same respective planes we plot also the coupling coefficients values corresponding to each viable data point, determined numerically from the Lagrangian without any approximation, for which we plot together the values of $a_u$ (gray dots) and $a_d$ (black dots) along the $a_f$ dimension. While $a_\Sigma$ and even more $a_u$ seem to be underconstrained by the current data, about half of the scanned data points stretch outside the 68\% CL region along the $a_S$ direction, and a few $a_d$ values lie outside the 99\% CL region. 

\begin{figure}[htb]
\centering
\includegraphics[width=0.46\textwidth]{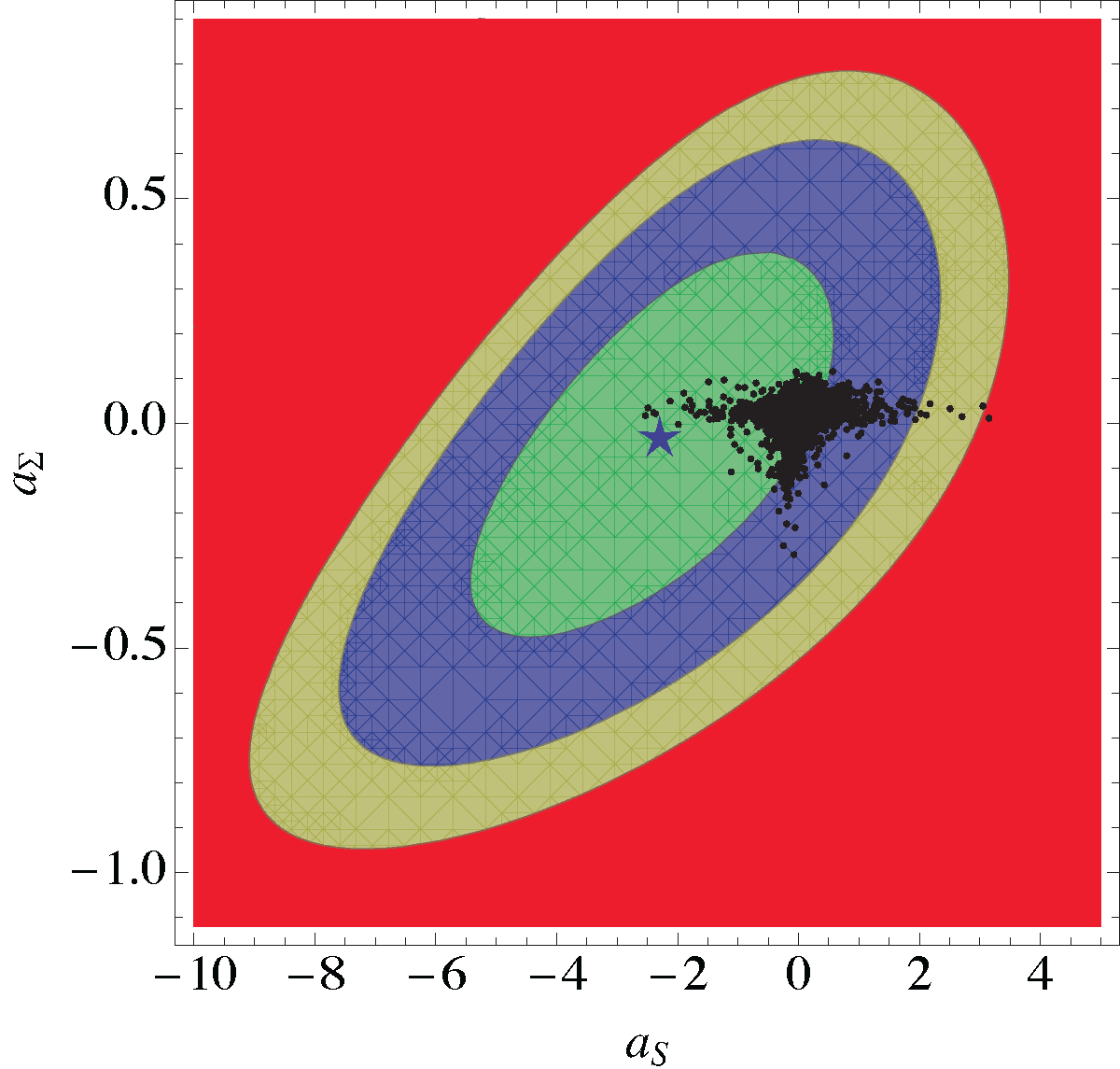}\hspace{0.1cm}
\caption{Viable regions at the 68\%, 95\%, 99\% CL in the coupling coefficients $a_S,a_\Sigma$ plane passing through the optimal point (blue star), together with the corresponding value (black) associated with each viable point.}
\label{aSaSigma}
\end{figure}

In Figs.~\ref{aSaSigma} we plot the 68\%, 95\%, 99\% CL viable regions (respectively in green, blue, and yellow) on the plane $a_S-a_\Sigma$ intersecting the optimal point (blue star) defined in Eq.~\eqref{optimusprime}, together with the corresponding coupling coefficients values for each viable data point (black).  No viable data point matches the optimal values, as the bulk of data points deviates from it about $1\sigma$ along the $a_S$ axis. While $a_S$ seems to be still underconstrained, we can expect the viable regions to shrink considerably with the next run of the LHC at 14 TeV, in which case the constraint on $a_S$ might become relevant if the optimal values do not change considerably.

\begin{figure}[htb]
\centering
\includegraphics[width=0.46\textwidth]{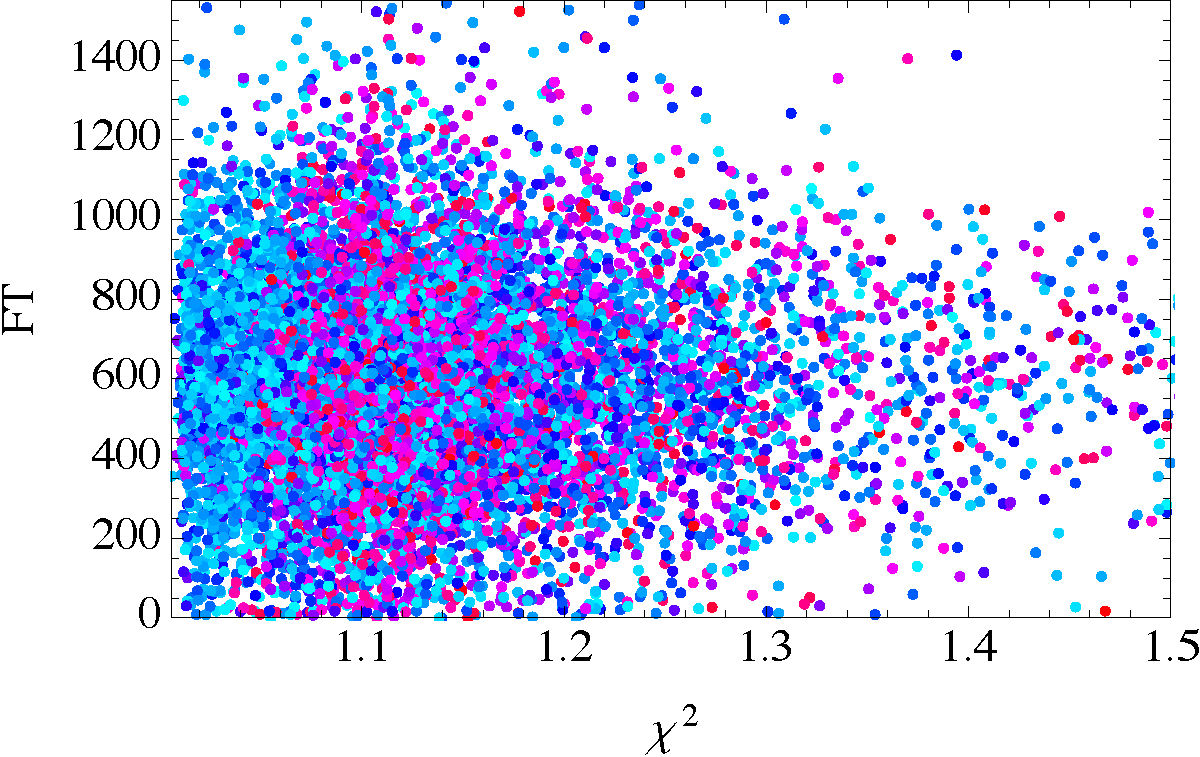}\hspace{0.1cm}
\includegraphics[width=0.12\textwidth]{lambdacscale.png}\hspace{0.1cm}
\caption{FT as a function of $\chi^2$ with colour code associated with the absolute value of $\lambda$. Mostly because of the deviation of the TESSM prediction on $\mathcal{B}r(B_s\to X_s \gamma)$ with the measured value the goodness of the fit worsens for points featuring large values of $\lambda$, which are also those that generally can achieve the smallest FT values.}
\label{chi2FT}
\end{figure}

Finally, in Fig.~\ref{chi2FT} we plot the FT for each data point, with the colour code of the absolute value of $\lambda$ defined in Fig.~\ref{lambdaFT2}, as a function of its $\chi^2$ value, which includes the contribution of $\mathcal{B}r(B_s\to X_s \gamma$) defined  in Eq.~\eqref{chi2}. As we can see from Fig.~\ref{bsgtan}, small $|\lambda|$ values more likely satisfy the $\mathcal{B}r(B_s\to X_s \gamma)$ experimental bound. It is important to notice that large absolute values of $\lambda$ are not able to improve the fit to current Higgs physics data enough to compensate for the bad fit to $\mathcal{B}r(B_s\to X_s \gamma$). The situation, though, has already changed considerably with the latest CMS data \cite{CMS:2014ega}, which has increased the significance of the enhancement of the Higgs decay to diphoton, favouring large $|\lambda|$ values. In a scenario in which both ATLAS and CMS confirm this enhancement with smaller uncertainty in the next LHC run, the TESSM would achieve a goodness of fit comparable to that of MSSM, with possibly a considerably smaller amount of FT.

\section{Conclusions}\label{concsec}
In this article we studied the phenomenology of the Triplet Extended Minimal Supersymmetric Standard Model, or TESSM, by first working out the neutral scalar masses at one loop using the Coleman-Weinberg potential and evaluating numerically the derivatives with respect to the neutral scalar fields. We performed a scan of the parameter space and found around 13000 points that satisfy direct search constraints besides producing the observed SM mass spectrum. Among these data points, we have shown that for large absolute values of the triplet coupling $\lambda$ it is possible to reach a smaller value of Fine-Tuning (FT) than for MSSM. Moreover, for large values of $|\lambda|$ it is possible to access regions of small $\tan\beta$ or/and small cubic stop coupling $A_t$, which are not accessible within MSSM with stop masses at the TeV scale. 

To check that the couplings remain perturbative at the given UV scale, which we chose to be equal to $10^4$ TeV, the highest scale tested through flavour observables, we calculated the full two-loop beta functions and required all the dimensionless couplings to be smaller than $2\pi$: some of the points which would be non-perturbative at one loop order indeed feature perturbativity at two loop order. 

To determine the phenomenological viability of TESSM we performed a goodness of fit analysis by comparing the TESSM predictions with 59 observables, comprising the $B_s\to X_s \gamma$ branching ratio, which we calculated at the next to leading order, as well as the light Higgs decays to $WW$, $ZZ$, $\tau\tau$, $b\bar{b}$, and all the topologies of $\gamma\gamma$, with experimental data from ATLAS, CMS, and Tevatron.  A new result we obtained is the possibility of a suppression of the Higgs decay to diphoton, generated mostly for values of $M_2$, the wino soft mass, with sign opposite to that of $\mu_D$, the superpotential mass of the two Higgs doublets. 

For large absolute values of $\lambda$ TESSM generates a large suppression or enhancement of the loop induced Higgs decay rate to diphoton. We find though that for large $|\lambda|$, or equivalently small $\tan\beta$, the values of $\mathcal{B}r(B_s\to X_s \gamma$) are always suppressed, with a deviation from the experimental value beyond $2\sigma$ for about half of the viable data points. The $\mathcal{B}r(B_s\to X_s \gamma$) values for small $|\lambda|$ instead feature both suppression and enhancement as compared with the measured value, with about half of the viable data points deviating less than $2\sigma$ from the experimental value. As a consequence, the goodness of fit of the 59 observables generally improves for smaller values of $|\lambda|$, for which the role of the triplet fields becomes less relevant in increasing the light Higgs mass and enhancing or suppressing the light Higgs decay to diphoton. The situation, though, has already changed considerably with the latest CMS data \cite{CMS:2014ega}, which has increased the significance of the enhancement of the Higgs decay to diphoton, favouring large $|\lambda|$ values. It is expected that the coming run of LHC will help the experiments to improve the accuracy of the Higgs branching ratios measurements. If the excess in the diphoton channel remains the same, the goodness of fit for TESSM would become comparable to that for MSSM, with FT in TESSM likely much smaller than in MSSM.

\section*{Acknowledgements}

The authors kindly thank Victor Martin-Lozano for useful discussions and for checking independently that a suppression of the Higgs decay rate indeed arises for suitable values of the free parameters. The authors acknowledge support from the Academy of Finland (Project Nro 137960). The work of ASK is also supported by the Finnish Cultural Foundation.

\appendix

\section{Appendix: Mass Matrices in TESSM}
\label{massmapp}
Here we list the field dependent mass matrices of TESSM where we keep only the real components of the neutral scalar fields. The reason behind this is that we consider only the 1-loop corrections to the CP even mass matrix calculated via the effective formula in Eq.~(\ref{1Lmh}). 

First, we present the mass matrices of the scalar sector of TESSM Lagrangian. When the CP symmetry is intact, the neutral Higgs mass matrices can be defined separately  for CP odd and even Higgses. The field dependent squared mass matrix for CP even Higgses in the basis $\frac{1}{\sqrt{2}}(a_u,a_d,a_T)$ is 
\begin{eqnarray}
\mathcal{M}^2_{h^0}= \begin{blockarray}{(c@{}c@{}c)}
\BAmulticolumn{2}{c}{\multirow{2}{*}{$ (\mathcal{M}^2_{h^0})_{2\times2}$}}& M_{13}^2\\
& & M_{23}^2\\
M_{13}^2 \;\;\;& M_{23}^2 \;\;\;& M_{33}^2\\
\end{blockarray}
\label{mns}
\end{eqnarray}
where 
\begin{eqnarray}
 &&(\mathcal{M}^2_{h^0_0})_{2\times2}=\nonumber
\\&&\resizebox{1.07\hsize}{!}{$\left(
\begin{array}{cc}
 m_{H_u}^2 + \frac{1}{4} \lambda ^2 a_d^2-\frac{1}{8}G^2(a_d^2-3a_u^2)+ \left( \mu _D-\frac{1}{2}\lambda  a_T\right){}^2 & -B_D \mu _D +\frac{ \lambda a_T}{2}(A_T+2 \mu_T) -\frac{1}{4}(G^2-2\lambda^2) a_d a_u\\
 -B_D \mu _D+ \frac{\lambda a_T}{2}(A_T+2 \mu _T)-\frac{1}{4}(G^2-2\lambda^2) a_d a_u& m_{H_d}^2+\frac{1}{4} \lambda ^2 a_u^2+\frac{1}{8}G^2 (3a_d^2-a_u^2)+\left( \mu _D-\frac{1}{2}\lambda  a_T\right){}^2 \\
\end{array}
\right),$}\nonumber\end{eqnarray}

\noindent
with $G^2=(g_Y^2+g_L^2)$ and
\begin{eqnarray}
M_{13}^2&=&\frac{1}{2} \lambda  \left(a_u \left(\lambda  a_T-2 \mu _D\right)+a_d \left(A_T+2 \mu _T\right)\right),\nonumber\\
M_{23}^2&=&\frac{1}{2} \lambda  \left(a_d \left(\lambda  a_T-2 \mu _D\right)+a_u \left(A_T+2 \mu _T\right)\right),\nonumber\\
M_{33}^2&=&m_T^2+\frac{1}{4} \lambda ^2 \left(a_d^2+a_u^2\right)+2 \mu _T \left(B_T+2 \mu _T\right).\nonumber
\end{eqnarray}
Similarly the mass squared matrix for CP odd Higgses can be written in the basis $\frac{1}{\sqrt{2}}(b_u,b_d,b_T)$ as  
\begin{eqnarray}
\mathcal{M}^2_{A}= \begin{blockarray}{(c@{}c@{}c)}
\BAmulticolumn{2}{c}{\multirow{2}{*}{ ($\mathcal{M}^2_{A})_{2\times2}$}}& M^{\prime 2}_{13}\\
& & M^{\prime 2}_{23}\\
M^{\prime 2}_{13} \;\;\;& M^{\prime 2}_{23} \;\;\;& M^{\prime 2}_{33}\\
\end{blockarray}\, ,
\end{eqnarray}
where
\begin{eqnarray}
 &&(\mathcal{M}^2_{A})_{2\times2}=\nonumber \\
&&\resizebox{1.07\hsize}{!}{$\left(
\begin{array}{cc}
m_{H_u}^2 +\frac{1}{4} \lambda ^2 a_d^2-\frac{1}{8}G^2 (a_d^2-a_u^2)+\left(\mu_D-\frac{1}{2}\lambda a_T \right){}^2 & B_D \mu _D-\frac{1}{2} \lambda  a_T \left(A_T+2 \mu _T\right) \\
 B_D \mu _D-\frac{1}{2} \lambda  a_T \left(A_T+2 \mu _T\right) & m_{H_d}^2+\frac{1}{4} \lambda ^2a_u^2+\frac{1}{8}G^2 \left(a_d^2-a_u^2\right)+\left(\mu_D-\frac{1}{2}\lambda a_T \right){}^2\\
\end{array}
\right)$}
\nonumber\end{eqnarray}

\noindent
and
\begin{eqnarray}
M_{13}^{\prime2}&=&-\frac{1}{2} \lambda  a_d \left(A_T-2 \mu _T\right),
\nonumber\\
M_{23}^{\prime2}&=&-\frac{1}{2} \lambda  a_u \left(A_T-2 \mu _T\right),\nonumber\\
M_{33}^{\prime2}&=&m_T^2+\frac{1}{4}\lambda ^2 \left( a_d^2+a_u^2\right)-2 B_T \mu _T+4 \mu _T^2.
\nonumber
\end{eqnarray}
The charged Higgs mass square matrix of TESSM in the basis $(H_u^+,H_d^{-*},T_2^+,T_1^{-*})$  can be written using $2\times 2$ submatrices as
\begin{eqnarray}
\mathlarger{ \mathcal{M}^2_{h^{\pm}}}=\left(
\begin{array}{cc}
(\mathcal{M}_{11}^2)_{2\times2}&(\mathcal{M}_{12}^2)_{2\times2}\\
(\mathcal{M}_{12}^2)^T_{2\times2}&(\mathcal{M}_{22}^2)_{2\times2}
\\
\end{array}\right)
 \end{eqnarray}
where 
\begin{eqnarray}
&&(\mathcal{M}_{11}^2)_{2\times2}=\nonumber\\
&&\resizebox{1.07\hsize}{!}{$\left(
\begin{array}{cc}
m_{H_u}^2+\frac{g_L^2 a_u^2}{4}-\frac{ \left(g_Y^2-g_L^2\right)}{8}
   \left(a_d^2-a_u^2\right)+\frac{a_d^2 \lambda ^2}{2}+\left(\frac{\lambda  a_T}{2}+\mu_D\right)^2 &  \frac{a_u a_d}{4}  \left(g_L^2+\lambda ^2\right)+\frac{\lambda  a_T}{2}(A_T+2\mu_T)+B_D \mu _D\\
  \frac{a_u a_d}{4}  \left(g_L^2+\lambda ^2\right)+\frac{1}{2}\lambda  a_T(A_T +2 \mu_T)+ B_D \mu _D & m_{H_d}^2+\frac{g_L^2 a_d^2}{4} +\frac{\left(g_Y^2-g_L^2\right)}{8}\left(a_d^2-a_u^2\right) +\frac{a_u^2 \lambda ^2}{2}+\left(\frac{a_T \lambda}{2}+\mu_D\right)^2 \\
\end{array}
\right)$}\,,\nonumber
 \end{eqnarray}

\begin{eqnarray}
&&(\mathcal{M}_{12}^2)_{2\times2}=\nonumber \\
&&\resizebox{1.07\hsize}{!}{$\left(
\begin{array}{cc}
\frac{1}{2\sqrt{2}}\left(\lambda ^2-g_L^2\right) a_T a_u+\frac{\lambda}{\sqrt{2}} \left(a_u \mu _D-2 a_d \mu_T\right) & \frac{ -A_T \lambda a_d}{\sqrt{2}} +\frac{a_u}{2\sqrt{2}}((-\lambda ^2+g_L^2) a_T+2\lambda \mu_D)\\
\frac{ A_T \lambda  a_u}{\sqrt{2}}+\frac{a_d}{2\sqrt{2}} \left(\left(\lambda ^2-g_L^2\right) a_T-2 \lambda  \mu _D\right)&\- \sqrt{2}\lambda  \mu_T a_u+\frac{1}{2\sqrt{2}}a_d\left(\left(-\lambda ^2+g_L^2\right) a_T-2 \lambda  \mu_D\right)\\
\end{array}
\right)$}\,,\nonumber
 \end{eqnarray}

\begin{eqnarray}
&&(\mathcal{M}_{22}^2)_{2\times2}=\nonumber \\
&&\resizebox{1.07\hsize}{!}{$\left(
\begin{array}{cc}
m_T^2+\frac{1}{2}\lambda ^2 a_u^2+\frac{1}{4} g_L^2 \left(a_d^2+2 a_T^2-a_u^2\right)+4 \mu _T^2 & -\frac{1}{2}g_L^2 a_T^2+2 B_T \mu _T\\
 -\frac{1}{2}g_L^2 a_T^2+2 B_T \mu _T & m_T^2+\frac{1}{2}\lambda ^2 a_d^2-\frac{1}{4} g_L^2 \left(a_d^2-a_u^2-2 a_T^2\right)+4 \mu _T^2\\
\end{array}
\right).$}\nonumber \end{eqnarray}

The second sector that we consider is the sfermion sector. The sfermion mass matrices of TESSM are slightly different from the MSSM ones because of an additional term appearing in the off-diagonal terms.  The mass matrices of stops and sbottoms written in the corresponding bases, $\left(\tilde{t}_L,\tilde{t}_R\right)$ and $\left(\tilde{b}_L,\tilde{b}_R\right)$, are 
respectively,
\begin{eqnarray}
\mathcal{M}_{\tilde{t}}=\left(
\begin{array}{cc}
m_{\tilde{Q}_{3L}}^2+\frac{1}{2} y_t^2 a_u^2-\frac{1}{24}\left(g_Y^2-3 g_L^2\right) \left(a_d^2-a_u^2\right) & \frac{y_t \left(2 A_t a_u+a_d \left(\lambda  a_T-2 \mu _D\right)\right)}{2 \sqrt{2}}\\
 \frac{y_t \left(2 A_t a_u+a_d \left(\lambda  a_T-2 \mu _D\right)\right)}{2 \sqrt{2}} &  m_{\tilde{t}_R}^2+\frac{1}{2} y_t^2
   a_u^2+\frac{g_Y^2}{6} \left(a_d^2-a_u^2\right)\\
\end{array}
\right),\end{eqnarray} 

\begin{eqnarray}
\mathcal{M}_{\tilde{b}}=
\left(
\begin{array}{cc}
 m_{\tilde{Q}_{3L}}^2+\frac{1}{2} y_b^2 a_d^2-\frac{1}{24}\left(g_Y^2+3 g_L^2\right)\left(a_d^2 - a_u^2\right) & \frac{y_b(2A_b a_d+ a_u
   \left(\lambda  a_T-2 \mu _D\right))}{2 \sqrt{2}} \\
 \frac{y_b(2 A_b a_d+a_u \left(\lambda  a_T-2 \mu _D\right))}{2 \sqrt{2}} & m_{\tilde{b}_R}^2+\frac{1}{2} y_b^2 a_d^2-\frac{g_Y^2}{12}\left(a_d^2-a_u^2\right) \\
\end{array}
\right).
\end{eqnarray}

Other contributions to the neutral Higgs mass at one loop come from neutralinos and charginos. Below we provide the field dependent mass matrix of neutralinos in the basis ($\tilde{B}^0$, $\tilde{W}^0$, $\tilde{H}_d^0$, $\tilde{H}_u^0$, $\tilde{T}^0$) as 
\begin{eqnarray}
\mathcal{M}_{\chi^0}=
\left(
\begin{array}{ccccc}
 M_1\;\; & 0\;\; & -\frac{1}{2} g_Y a_d\;\; & \frac{1}{2} g_Y a_u\;\; & 0 \\
 0\;\; & M_2\;\; & \frac{1}{2} g_L a_d\;\; & -\frac{1}{2} g_L a_u\;\; & 0 \\
 -\frac{1}{2} g_Y a_d\;\; & \frac{1}{2} g_L a_d\;\; & 0\;\; & -\mu _D+\frac{1}{2}\lambda  a_T \;\;&
   \frac{1}{2} \lambda  a_u\\
 \frac{1}{2} g_Y a_u\;\; & -\frac{1}{2} g_L a_u\;\; & -\mu _D+\frac{1}{2} \lambda  a_T\;\; & 0\;\; &
   \frac{1}{2} \lambda  a_d \\
 0\;\; & 0\;\; & \frac{1}{2} \lambda  a_u\;\; & \frac{1}{2}\lambda  a_d\;\; & 2 \mu_T \\
\end{array}
\right).
\end{eqnarray}

\noindent
For the chargino sector, the mass matrix appears in the Lagrangian with the three column vectors $\psi^+=$($\tilde{W}^+$,$\tilde{H}_u^+$,$\tilde{T}_2^+$) and $\psi^-=$($\tilde{W}^-$,$\tilde{H}_d^-$,$\tilde{T}_1^-$) as 
\begin{eqnarray}
\mathcal{L}\supset - (\psi^-)^{T} \mathcal{M_{\chi}} \psi^+ + h.c\ ,
\end{eqnarray}
where 
\begin{eqnarray}
M_{\chi^{\pm}}=\left(
\begin{array}{ccc}
 M_2\;\; & \frac{1}{\sqrt{2}}g_L a_u\;\; & -g_L a_T \\
 \frac{1}{\sqrt{2}}g_L a_d\;\; & \frac{1}{2} \lambda  a_T+\mu _D\;\; & \frac{1}{\sqrt{2}}\lambda  a_u
   \\
 g_L a_T\;\; & \frac{-1}{\sqrt{2}}\lambda  a_d\;\; & 2 \mu _{T} \\
\end{array}
\right).  
\end{eqnarray}

\section{Beta Functions at 2 Loops} \label{betas}
We assume the first two families to have negligible Yukawa couplings. The only dimensionless couplings are therefore the Yukawa couplings $y_t,y_b,y_\tau,\lambda$, as well as the gauge couplings $g_3,g_2=g_L,g_1=\sqrt{5/3}\,g_Y$, for which the beta functions at two loops \cite{Martin:1993zk} are defined by
\be
\frac{d z_x}{d t}=\frac{\beta _x^{\text{(1)}}}{16 \pi ^2}+\frac{\beta _x^{\text{(2)}}}{\left(16 \pi ^2\right)^2}\quad , \quad z_x=g_1,g_2,g_3,y_t,y_b,y_\tau,\lambda_T\quad ,\quad t=\log\frac{E}{E_0}\ ,
\ee
with $\lambda_T=\lambda$. We find, in the renormalization scheme using dimensional reduction (see \cite{Martin:1997ns} and references therein) with modified minimal subtraction ($\overline{\rm DR}$):
\bea
\beta _1^{\text{(1)}}&=&\frac{33 g_1^3}{5}\ ,
\nonumber\\ 
\beta _1^{\text{(2)}}&=&-\frac{9}{5} \lambda ^2 g_1^3+\frac{199 g_1^5}{25}+\frac{27}{5} g_1^3 g_2^2+\frac{88}{5} g_1^3 g_3^2-\frac{14}{5} g_1^3 y_b^2-\frac{26}{5} g_1^3
   y_t^2-\frac{18}{5} g_1^3 y_{\tau }^2\ ,
\nonumber\\ 
\beta _2^{\text{(1)}}&=&3 g_2^3\ ,
\nonumber\\ 
\beta _2^{\text{(2)}}&=&-7 \lambda ^2 g_2^3+\frac{9}{5} g_1^2 g_2^3+49 g_2^5+24 g_2^3 g_3^2-6 g_2^3 y_b^2-6 g_2^3 y_t^2-2 g_2^3 y_{\tau }^2\ ,
\nonumber\\ 
\beta _3^{\text{(1)}}&=&-3 g_3^3\ ,
\nonumber\\ 
\beta _3^{\text{(2)}}&=&\frac{11}{5} g_1^2 g_3^3+9 g_2^2 g_3^3+14 g_3^5-4 g_3^3 y_b^2-4 g_3^3 y_t^2\ ,
\nonumber\\
\beta _t^{\text{(1)}}&=&y_t \left(\frac{3 \lambda ^2}{2}-\frac{13 g_1^2}{15}-3 g_2^2-\frac{16 g_3^2}{3}+y_b^2+6 y_t^2\right)\ ,
\nonumber\\ 
\beta _t^{\text{(2)}}&=&y_t \left(-\frac{15 \lambda^4}{4}+\frac{2743 g_1^4}{450}+6 \lambda ^2 g_2^2+g_1^2 g_2^2+\frac{27 g_2^4}{2}+\frac{136}{45} g_1^2 g_3^2+8 g_2^2
   g_3^2-\frac{16 g_3^4}{9}-6 \lambda ^2 y_b^2\right.\nonumber\\&+&\left.\vphantom{\frac{2}{5}} g_1^2 y_b^2-5 y_b^4-\frac{9}{2} \lambda ^2 y_t^2+\frac{6}{5} g_1^2 y_t^2+6 g_2^2 y_t^2+16 g_3^2 y_t^2-5 y_b^2 y_t^2-22
   y_t^4-\frac{3}{2} \lambda ^2 y_{\tau }^2-y_b^2 y_{\tau }^2\right)\ ,
\nonumber\\ 
\beta _b^{\text{(1)}}&=&y_b \left(\frac{3 \lambda ^2}{2}-\frac{7 g_1^2}{15}-3 g_2^2-\frac{16 g_3^2}{3}+6 y_b^2+y_t^2+y_{\tau }^2\right)\ ,
\nonumber\\ 
\beta _b^{\text{(2)}}&=&y_b \left(-\frac{15 \lambda^4}{4}+\frac{287 g_1^4}{90}+6 \lambda ^2 g_2^2+g_1^2 g_2^2+\frac{27 g_2^4}{2}+\frac{8}{9} g_1^2 g_3^2+8 g_2^2
   g_3^2-\frac{16 g_3^4}{9}-\frac{9}{2} \lambda ^2 y_b^2\right.\nonumber\\&+&\left.\vphantom{\frac{2}{5}} g_1^2 y_b^2+6 g_2^2 y_b^2+16 g_3^2 y_b^2-22 y_b^4-6 \lambda ^2 y_t^2+\frac{4}{5} g_1^2 y_t^2-5 y_b^2 y_t^2-5
   y_t^4+\frac{6}{5} g_1^2 y_{\tau }^2-3 y_b^2 y_{\tau }^2-3 y_{\tau }^4\right)\ ,
\nonumber\\
\beta _{\tau }^{\text{(1)}}&=&y_{\tau } \left(\frac{3 \lambda ^2}{2}-\frac{9 g_1^2}{5}-3 g_2^2+3 y_b^2+4 y_{\tau }^2\right)\ ,
\nonumber\\ 
\beta _{\tau }^{\text{(2)}}&=&y_{\tau } \left(-\frac{15 \lambda^4}{4}+\frac{27 g_1^4}{2}+6 \lambda ^2 g_2^2+\frac{9}{5} g_1^2 g_2^2+\frac{27 g_2^4}{2}-\frac{2}{5} g_1^2
   y_b^2+16 g_3^2 y_b^2-9 y_b^4-\frac{9}{2} \lambda ^2 y_t^2\right.\nonumber\\&-&3 y_b^2 y_t^2-\left.\vphantom{\frac{9}{2}} \lambda ^2 y_{\tau }^2+\frac{6}{5} g_1^2 y_{\tau }^2+6 g_2^2 y_{\tau }^2-9 y_b^2 y_{\tau }^2-10
   y_{\tau }^4\right)\ ,
 \nonumber\\
 \beta _T^{\text{(1)}}&=&\lambda  \left(4 \lambda ^2-\frac{3 g_1^2}{5}-7 g_2^2+3 y_b^2+3 y_t^2+y_{\tau }^2\right)\ ,
\nonumber\\ 
 \beta _T^{\text{(2)}}&=&\lambda  \left(-\frac{21 \lambda^4}{2}+\frac{3}{5} \lambda ^2 g_1^2+\frac{207 g_1^4}{50}+11 \lambda ^2 g_2^2+\frac{9}{5} g_1^2 g_2^2+\frac{83
   g_2^4}{2}-\frac{15}{2} \lambda ^2 y_b^2-\frac{2}{5} g_1^2 y_b^2+16 g_3^2 y_b^2\right.\nonumber\\&-&9 y_b^4-\left.\vphantom{\frac{15}{2}} \lambda ^2 y_t^2+\frac{4}{5} g_1^2 y_t^2+16 g_3^2 y_t^2-6 y_b^2 y_t^2-9
   y_t^4-\frac{5}{2} \lambda ^2 y_{\tau }^2+\frac{6}{5} g_1^2 y_{\tau }^2-3 y_{\tau }^4\right)\ .
\eea



\begin{thebibliography}{99}
\bibitem{CMS:2014ega} 
  CMS Collaboration [CMS Collaboration],
and studies of the compatibility of its couplings with the standard model,''
  CMS-PAS-HIG-14-009.
\end{thebibliography}
\end{document}